\documentclass[useAMS,usenatbib]{mn2e}
\usepackage{journal_abbreviations}
\usepackage{graphicx, amsmath, amssymb}
\usepackage{enumerate}
\usepackage{float}
\usepackage[T1]{fontenc}
\usepackage{aecompl}

\usepackage{mathrsfs}
\usepackage{graphics}

\title[Implications of a variable IMF]{Implications of a variable IMF for the interpretation of observations of galaxy populations}
\author[Clauwens et al.]{Bart Clauwens$^{1,2}$\thanks{E-mail: clauwens@strw.leidenuniv.nl}, Joop Schaye$^{1}$, Marijn Franx$^{1}$\\
$^{1}$Leiden Observatory, Leiden University, PO Box 9513, 2300 RA Leiden, The Netherlands\\$^{2}$Instituut-Lorentz for Theoretical Physics, Leiden University, 2333 CA Leiden, The Netherlands\\}

\begin{document}

\date{Accepted 2016 July 21. Received 2016 July 21; in original form 2016 March 16.}

\pagerange{\pageref{firstpage}--\pageref{lastpage}}

\maketitle

\label{firstpage}

\begin{abstract}

We investigate the effect of a metallicity-dependent stellar initial mass function (IMF), as deduced observationally by \citet{MartinNavarro15}, on the inferred stellar masses and star formation rates (SFRs) of a representative sample of 186,886 SDSS galaxies. Relative to a Chabrier IMF, for which we show the implied masses to be close to minimal, the inferred masses increase in both the low- and
high-metallicity regimes due to the addition of stellar remnants and dwarf stars, respectively. The resulting galaxy stellar mass function (GSMF) shifts toward higher masses by 0.5 dex, without affecting the high-mass slope (and thus the need for effective quenching). The implied low-redshift SFR density increases by an order of magnitude. However, these results depend strongly on the assumed IMF parametrisation, which is not directly constrained by the observations. Varying the low-end IMF slope instead of the high-end IMF slope, while maintaining the same dwarf-to-giant ratio, results in a much more modest GSMF shift of 0.2 dex and a 10 per cent increase in the SFR density relative to the Chabrier IMF. A bottom-heavy IMF during the late, metal-rich evolutionary stage of a galaxy would help explain the rapid quenching and the bimodality in the galaxy population by on the one hand making galaxies less quenched (due to the continued formation of dwarf stars) and on the other hand reducing the gas consumption timescale. We conclude that the implications of the observational evidence for a variable IMF could vary from absolutely dramatic to mild but significant.

\end{abstract}


\section{Introduction}
\label{SectionIntroduction}

The stellar initial mass function (IMF) describes the number distribution of stars as a function of stellar mass at the zero age main sequence. The IMF can not only help us understand physical processes inside molecular clouds, it is also a key ingredient for understanding astrophysical processes on larger scales. A top-heavy IMF will lead to more energetic stellar feedback, a higher metal production and more ionizing radiation, which could have an impact on the nature of the formed galaxies, the total star formation history, metal enrichment and on the re-ionization history of our Universe.

Although IMF studies have a rich history going back to the seminal work of \citet{Salpeter55}, both in nearby galaxies through direct star counts and in more distant galaxies by more indirect methods, a consensus on the functional form of the IMF and its possible dependencies has not been reached. Observations aiming to measure galaxy stellar masses and star formation rates typically assume a \citet{Chabrier03} IMF \citep[e.g.][]{Moustakas13,Bauer13,Chang15} or a similar \citet{Kroupa01} IMF \citep[e.g.][]{Kauffmann03,Muzzin13}. The same holds true for hydrodynamic simulations of galaxy formation \citep[e.g.][]{Hopkins14,Schaye15,Snyder15}. A different IMF, specifically an IMF that depends on one or more local astrophysical conditions, could have a dramatic impact on the interpretation of observations and the results of simulations.

Most nearby IMF measurements tend to support a universal Kroupa- or Chabrier-like IMF, see \citet{Kroupa02,Chabrier03,Kroupa13,Kirk11,Bastian11} and references therein, but there are many complications in determining the IMF from direct star counts, see e.g. \citet{Scalo05} and \citet{Weisz13}. Recently, \citet{Weisz15} found a very small intrinsic scatter in the power-law slope of the high-end mass function for young resolved star clusters in M31, albeit at a higher value (1.45$\pm$0.05) than the Kroupa slope (1.30) or the slope they infer for the Milky Way (1.15$\pm$0.1), although this offset might be explained by dynamical depletion \citep{Oh16}. Conversely, a recent dynamical study of 29 Local Group Clusters by \cite{Zaritsky14} concludes that there are clear variations in the cluster IMF. Indications for a nearby metallicity-dependent IMF come from stellar counts of two nearby metal-poor, old, ultra-faint dwarf galaxies by \citet{Geha13}, which are found to have a very top-heavy IMF, and from a study of galactic globular clusters by \citet{Marks12}, who report an increasingly top-heavy IMF with decreasing cluster metallicity and increasing cloud-core density.

IMF studies based on observations of more distant galaxies can be roughly divided into two classes. The first class consists of dynamical methods that weigh the stellar content of a galaxy and compare this to the stellar mass obtained from spectral energy density (SED) fitting. Studies of the dynamical masses of SDSS galaxies by \citet{Dutton13,Conroy12,Tortora13} report IMF trends with velocity dispersion or galaxy mass. However, a similar study of low-mass (around $\rm{10^{9} M_{\odot}}$) early-type galaxies (ETGs) by \citet{Tortora15} reports the same average IMF normalisation for these low-mass ETGs as for the more massive ETGs. All these studies report relatively large variations in the IMF mass normalisation at fixed stellar mass, but it is not exactly clear to what extent these variations could be caused by assumptions on the dark matter profile, the anisotropy of stellar motions or the inclination of the observed galaxies. A major risk is confusion of dark matter and stellar matter. For this reason the centres of ETGs are the best targets as they are believed to be relatively devoid of dark matter.

The most ambitious attempt at measuring dynamical stellar masses of ETG centres has been undertaken by the ATLAS$^{\rm{3D}}$ team, who use integral field spectroscopy of 260 nearby ETGs. \citet{Cappellari12,ATLAS15,ATLAS20} report IMF variations that correlate with the stellar mass-to-light ratio inferred from SED modelling under the assumption of a universal IMF and also with the effective velocity dispersion of the ETGs, but not with age, metallicity or alpha-enhancement \citep{McDermid14}. However, \citet{Clauwens15} show that the inferred IMF variations could be caused by a slight underestimation of the modeling- and measurement errors. Random Gaussian errors in the total kinematic mass determination of the order 30\% would give a trend that is very similar to the one reported in \citet{Cappellari12}. Recently, \citet{Li15} tested the accuracy of stellar mass measurements with the JAM method used by \citet{Cappellari12}, by applying it to mock galaxies from the Illustris simulation and they report random modeling errors of 40\% and biases that grow with diminishing resolution\footnote{This 40\% error is for the high-resolution test, which still has a lower resolution than is typical for ATLAS$^{\rm{3D}}$ observations. A similar analysis of mock galaxies at a higher resolution would be needed to disentangle the contributions of IMF variations and modelling- and measurement-errors to the observed variations in $M_{*kin}/M_{*Salp}$ found by the ATLAS$^{\rm{3D}}$ Survey. A resolution-dependent bias could also possibly cause a trend of $M_{*kin}/M_{*Salp}$ with distance, like the one reported in \citet{Clauwens15}.}.

Strong gravitational lensing provides an independent way to measure dynamical stellar masses. In general such studies find a Kroupa-like IMF normalization for spiral galaxies \citep{Brewer12} and a normalization heavier than Kroupa for massive ETGs \citep{Treu10,Oguri14,Barnabe13}, but the analysis of the nearest strong-lensing galaxy by \citet{Smith13}  does not support this\footnote{The strong-lensing measurement of this particular galaxy is expected to be less sensitive to dark matter, since the Einstein radius is roughly 25\% of the effective radius, probing a region that is believed to be dominated by stars.}.

The second class of independent methods to determine the galaxy scale IMF relies on the observation of specific spectral features that discriminate between low- and high-mass stars. A comparison of the X-ray binary count studies of \citet{Dabringhausen12} and \citet{Peacock14} suggests a top-heavy IMF for ultracompact dwarf galaxies. Several spectral studies point at an enhanced fraction of low-mass stars in ETGs with respect to a standard Kroupa IMF. \citet{Dokkum12} and \citet{Conroy12} find an IMF trend with alpha-enhancement, velocity dispersion and metallicity (in decreasing order of significance), compatible with results from \citet{Pastorello14}. Other studies \citep{Barbera13,MartinNavarro15c,MartinNavarro15b} suggest a trend that is driven by velocity dispersion or stellar mass. \citet{Smith14} and \citet{Smith15} show that there is disagreement between IMF results from spectral features and respectively kinematic- and strong lensing studies of the same galaxies.

Overall, IMF observations have not yet converged to a single view. The primary candidates for variables that correlate with IMF variations are metallicity and velocity dispersion. Some other candidates suggested by observations are dynamically determined density\citep{Spiniello15}, star formation rate \citep{Gunawardhana11}, luminosity \citep{Hoversten08}, surface brightness (\citeauthor{Meurer09} \citeyear{Meurer09} and \citeauthor{Lee09} \citeyear{Lee09}, but see \citeauthor{Boselli09} \citeyear{Boselli09}) and cosmic time \citep{Dave08,Ferraras15}. Of course all these variables are also correlated with each other in a complicated way. An ambitious attempt at resolving these interdependencies is the IGIMF theory \citep{Weidner13,Recchi15} which builds an effective galactic scale IMF in a bottom-up manner from the IMFs of individual star clusters, that depend on metallicity, SFR and cloud core density. In this paper we will take a more agnostic top-down approach and consider IMF variations on a galactic scale.

Distilling a consistent view from the reported constraints on the IMF is difficult, since the observations are performed on different types of galaxies with different diagnostics, different systematics, different model dependencies, different assumptions, a sensitivity to different mass ranges of the IMF and even different IMF parametrisations. For example, some spectroscopic studies use the so-called bimodal IMF parametrisation from \citet{Vazdekis96}, for which the free parameter is the IMF slope above 0.6 $\rm{M_{\odot}}$, whereas others use a parametrisation for which the IMF slope above 0.5 $\rm{M_{\odot}}$ or 1.0 $\rm{M_{\odot}}$ is kept fixed \citep{Conroy12}. In addition, sometimes the low-mass cut-off is allowed to vary \citep{Spiniello15}.

Still, the implications of IMF variations on derived galaxy properties could be large and are worth investigating. Previous work on the effect of IMF variations on inferred galaxy masses has focused specifically on the implications of an IMF trend with velocity dispersion. \citet{McGee14} report a significantly shallower galaxy stellar mass function (GSMF) high-mass drop-off based on IMF-velocity dispersion trends observed by \citet{Barbera13,Ferreras13,Spiniello14}. Conversely, \citet{Clauwens15} report a nearly unchanged GSMF high-mass drop-off, based on the IMF observations by \citet{Cappellari12}.

In this paper we investigate the effects of the IMF trend with metallicity observed by \citet{MartinNavarro15}, although many of the main conclusions also hold in a broader context. We will not only focus on the effects on inferred galaxy masses, but also on the inferred star formation rates, which turn out to be very sensitive to the assumed IMF.

\citet{MartinNavarro15} find a strong IMF dependency on metallicity, with high-metallicity regions containing more low-mass stars. They analyse integral field spectroscopic data from the CALIFA survey \citep{Sanchez12} for 24 ETGs in order to obtain radial IMF profiles, using the spectral indices $\rm{H_{\beta O}}$ , [MgFe]$^\prime$ , Mg2Fe, NaD, TiO$\rm{_{2 CALIFA}}$ and TiO$_1$.  A consistent IMF-metallicity trend is obtained for SDSS galaxy spectra that include near-IR IMF-sensitive features. Their sample shows a weaker IMF dependence on respectively age, velocity dispersion and [Mg/Fe]. The reported stronger trend with metallicity (Spearman rank correlation 0.82) rather than with [Mg/Fe] (Spearman rank correlation 0.21) is not necessarily in conflict with the results from \citet{Conroy12} who find Spearman rank correlations of respectively 0.33 with metallicity and 0.81 with [Mg/Fe], because the sample of \citet{MartinNavarro15} has a much wider range of metallicities, going as low as [M/H]$\rm{\approx-0.4}$.

In this paper we investigate the implications that this strong metallicity dependence of the IMF, concretely equation 2 from \citet{MartinNavarro15}, would have on several galaxy diagnostics. Specifically, we concentrate on the implications for galaxy stellar masses, star formation rates and derived relations like the galaxy stellar mass function, the main sequence of star formation and the mass-metallicity relation. We do this by re-interpreting the observations of a representative sample of 186,886 SDSS redshift $z\sim0.1$ galaxies. Although the presented results will be fully quantitative, our results should be viewed in a more qualitative way, due to the many uncertainties in measuring the IMF and the resulting diversity of the reported IMF measurements that have been discussed.

Although the physical mechanisms responsible for the formation of the IMF are not yet understood, metallicity, being a local property of the interstellar medium with a strong influence on the cooling rate of the gas, is likely to have an influence on the mass function of stars. On the other hand, a locally metallicity-dependent IMF could have a large influence on galaxies as a whole. A reduction in the formation of high-mass stars with respect to low-mass stars in the late, high-metallicity, stage of a galaxy's evolution could have an accelerating effect on the quenching of star formation in the galaxy, due to less energetic stellar feedback. This could trigger a last burst of low-mass star formation that empties the cool gas reservoir more rapidly. The true star formation rate of the galaxy would be larger than the typically inferred SFR, which is only sensitive to high-mass star formation.

The remainder of this paper is organised as follows. Section \ref{SectionMethod} describes the method used to translate the metallicity-dependent IMF variations into stellar mass and SFR changes for a large sample of SDSS galaxies. The implications for the star formation main sequence, GSMF and the mass-metallicity relation are presented in sections \ref{SectionMainSequence}, \ref{SectionGSMF} and \ref{SectionMetals}, respectively. Section \ref{SectionQuenching} discusses the possible implications of a IMF-metallicity trend on the quenching of galaxies. For an overview of the main takeaways of this work, see section \ref{SectionConclusions}. Some limitations of our analysis are discussed in the Appendix. Throughout this work we adopt a solar abundance of $\rm{Z_{\odot}= 0.019}$.

\section{Method}
\label{SectionMethod}

\subsection{The Vazdekis IMF}

\begin{figure}
\center
\includegraphics[width=1.0\columnwidth]{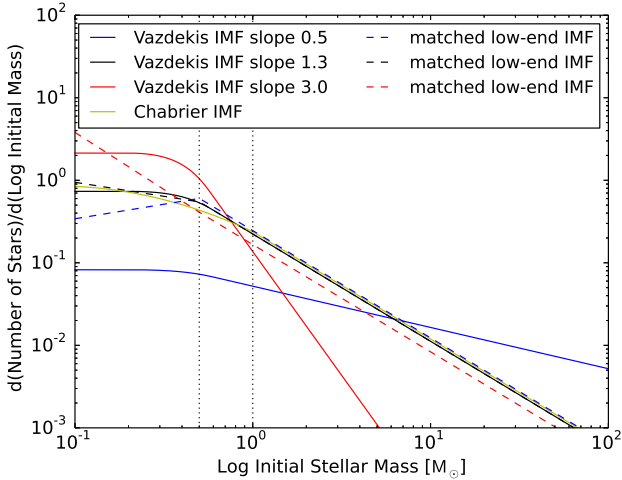}
\caption{The IMF for the two parametrisations used throughout this work. The yellow solid line depicts the Chabrier IMF. The other solid lines are for the Vazdekis IMF \protect{\citep{Vazdekis96}}, which is parametrised by its high-end slope. The indicated IMF slopes for the blue-, black- and red curves correspond to a $[M/H]$ value in equation \ref{eqone} of respectively -0.55, -0.29 and 0.26. Dashed lines show our corresponding low-end IMF, for which the slope below 0.5 $\rm{M_{\odot}}$ is the free parameter. Lines with the same colour are matched on the dwarf-to-giant ratio, equation \ref{eqfhalfone}, defined as the ratio of the mass in stars below 0.5 $\rm{M_{\odot}}$ with respect to the mass in stars below 1.0 $\rm{M_{\odot}}$. These mass regimes are indicated with the vertical dotted lines. The same dwarf-to-giant ratio, as measured from IMF sensitive spectral lines \protect{\citep{MartinNavarro15}}, can be accommodated by vastly different IMFs.}
\label{FigureParametrisation}
\end{figure}

The IMF can be uniquely defined by its slope as a function of stellar mass. The IMF slope is defined as the slope in the plane of (log number of stars per log initial stellar mass) versus (log initial stellar mass), see Figure \ref{FigureParametrisation}. It is customary to define this IMF slope to be positive for a descending curve.

\citet{MartinNavarro15} report a strong trend of the IMF slope with metallicity, expressed in their equation 2:
\begin{equation}
\Gamma_b  = 2.2(\pm0.1) + 3.1(\pm0.5) \times [M/H] ,
\label{eqone}
\end{equation}
where $[M/H]$ is the logarithm of the metallicity with respect to solar\footnote{Note that the inferred  high-end IMF slope is quite high for solar metallicity ($\Gamma_b=2.2$ for $Z=0.019$) which is possibly in tension with local IMF measurements, although adopting a solar metallicity of $Z=0.012$ lowers the inferred IMF slope to $\Gamma_b=1.58$. $\Gamma_b=1.3$ is reached for $Z=0.010$.} and  $\Gamma_b$ is the IMF slope above a stellar mass of 0.6 $\rm{M_{\odot}}$ in the bimodal IMF parametrisation of \citet{Vazdekis96}\footnote{This high-end IMF slope is defined such that the Salpeter value is 1.35 (and not 2.35).}. In this work we will refer to this IMF parametrisation as the Vazdekis IMF. We show some examples of this IMF for different values of $\Gamma_b$ in Figure \ref{FigureParametrisation}. This Vazdekis IMF has a slope of zero below a stellar mass of 0.2 and a spline from 0.2 $\rm{M_{\odot}}$ to 0.6 $\rm{M_{\odot}}$. The spline is normalized such that the endpoints would be the same if the IMF of the two outer parts would be interpolated and matched at 0.4 $\rm{M_{\odot}}$. The IMF is defined between 0.1 $\rm{M_{\odot}}$ and 100 $\rm{M_{\odot}}$. Equation \ref{eqone} is a fit to spectroscopic IMF determinations from local CALIFA measurements, SDSS measurements and radial profile measurements of three ETGs by \citet{MartinNavarro15}. These three data sets use a similar spectroscopic IMF analysis on different sources and give consistent results, as can be seen from fig. 2 of \citet{MartinNavarro15}. 

The aim of this paper is to apply this relation on a galaxy-by-galaxy basis to a large sample of SDSS galaxies and to investigate the potential consequences of such a metallicity-dependent IMF. We use the stellar metallicity measurements of \citet{Gallazzi05} as input for the metallicity-dependent IMF. In doing so we implicitly assume that these metallicity measurements themselves are not dependent on the IMF. This will not be completely accurate, but it is a valid approximation, since the typically IMF-dependent metal line indices TiO and NaD are excluded from the metallicity measurement of \citet{Gallazzi05}. They concentrate on those indices that are best reproduced by the \citet{Bruzual03} models with a fixed IMF\footnote{The metallicities measured by \citet{MartinNavarro15} are expected to deviate slightly from those of \citet{Gallazzi05}. Generally the age-sensitive features (e.g. H$\beta$)  have a slight sensitivity to the IMF \citep{Barbera13}, which results in slightly younger ages and thus slightly higher metallicities for a bottom-heavy IMF with respect to a Chabrier IMF.}. A recent mass-metallicity relation measurement by \citet{GonzalezDelgado14} with the CALIFA integral field spectrograph for a sample of 300 galaxies, ranging from ellipticals to late-type spirals, agrees well with the SDSS measurements by \citet{Gallazzi05}. At a given mass, the scatter in the metallicity is expected to be mostly intrinsic, rather than due to measurement uncertainties (see fig. 8 of \citeauthor{Gallazzi05} \citeyear{Gallazzi05}).

The spectroscopic IMF determination of \citet{MartinNavarro15} relies on the light being dominated by old stars. Therefore these measurements are done exclusively for ETGs. We will assume that the same relation holds also for star forming galaxies. Of course there is a possibility that a metallicity-IMF relation will manifest itself differently in star forming galaxies, but large differences are not expected, since the IMF is literally a property of the galaxy during star formation. Hence, for any given ETG, the measured IMF corresponds to the time when most of its stellar mass was build up, at which time it would be classified as a star forming galaxy. There is a possibility of a redshift dependence of equation \ref{eqone}, which could mean that it would be different for ETGs and star forming galaxies observed at redshift zero, but here we will assume that the IMF is solely determined by the local gas metallicity during star formation and does not depend directly on redshift. The measured stellar metallicity reflects this local gas metallicity during star formation.

Another simplifying assumption of our analysis is that we neglect metallicity gradients and characterise each galaxy by a single metallicity. This is justified by the fact that the IMF determinations of galaxy regions by \citet{MartinNavarro15} lie on the same metallicity-dependent trend as that of complete SDSS galaxies. Another caveat is that the CALIFA ETG sample consists mostly of high-velocity dispersion galaxies. We assume that the reported IMF-metallicity trend for this sample is representative for all galaxies. We will see that the implications of this IMF trend for inferred galaxy properties are most prominent at high masses and therefore are not expected to be affected much by this assumption.

The two main galaxy properties that we want to adjust to a metallicity-dependent IMF are galaxy stellar mass and star formation rate. We use the SDSS masses and star formation rates from \citet{Chang15} and the r-band weighted ages from \citet{Gallazzi05} . We match on Plate ID, Fiber ID and modified julian date and select the galaxies with a well-defined metallicity and age in \citet{Gallazzi05}. This leaves a sample of 186,886 SDSS galaxies.

To recompute the galaxy masses, for each galaxy we compare the Cousins R-band mass-to-light ratio of a simple stellar population (SSP) with the given metallicity and age and a Chabrier IMF\footnote{The Chabrier IMF \citep{Chabrier03} is defined as $\frac{d \,N}{d \, log_{10} (M/{\rm M_{\odot}})}=A_1 \, exp [-\frac{(log_{10}(M/{\rm M_{\odot}})-log_{10}(0.079))^2}{2\times 0.69^2}]$ for $M<1.0\, {\rm M_{\odot}}$ and $\frac{d \,N}{d \, log_{10} (M/{\rm M_{\odot}})}=A_2 \, (M/{\rm M_{\odot}})^{-1.3}$ for $M>1.0\, {\rm M_{\odot}}$, with $N$ the number of stars and $M$ the initial stellar mass. The normalisation constants $A_1$ and $A_2$ are fixed by the requirement that the IMF is continuous at $M=1.0\, {\rm M_{\odot}}$ and that the total integral of initial mass in stars from ${\rm 0.1 \,M_{\odot}}$ to ${\rm 100\, M_{\odot}}$ equals ${\rm 1 \,M_{\odot}}$. This IMF is very similar to the  \citep{Kroupa01} IMF and the "Canonical IMF" \citep{Kroupa13}.} to the mass-to-light ratio of a SSP with the same metallicity and age, but with the IMF implied by equation \ref{eqone}. We verified that the results are almost the same if we use the Johnson V-band or the SDSS r-band.

All results in this paper are obtained with the flexible stellar population synthesis (FSPS) code of \citet{Conroy10}, but where possible we verified that very similar results are obtained by using the tables from the MILES website. Metallicities are restricted to the range for which there are BaSTI isochrones \citep{Pietrinferni13}, $-1.80 < log_{10}(Z/{\rm Z_{\odot}}) < 0.32$, for a solar metallicity of $\rm{Z_{\odot}= 0.019}$ (which is used throughout this paper). The FSPS spectra are generated for IMF slope values binned with a separation of $\delta \Gamma =0.1$. Then the $M/L_R$ values are interpolated in the 3-dimensional space of (log age, metallicity, IMF slope).

We apply equation \ref{eqone} slightly outside the metallicity regime of the observations by \citet{MartinNavarro15} ($-0.40<log_{10}(Z/{ \rm Z_{\odot}})<0.25$). The highest metallicity in \citet{Gallazzi05} is $log_{10}(Z/{\rm Z_{\odot}})=0.417$. At the low-metallicity-end we extrapolate to $log_{10}(Z/{\rm Z_{\odot}})=-0.60$ or equivalently a Vazdekis IMF slope of 0.34, and keep the IMF constant for lower metallicities. This extrapolation, necessitated by a lack of measurements for galaxies in the metallicity range $-1.80 < log_{10}(Z/{\rm Z_{\odot}}) < -0.40$,  introduces a large uncertainty in the predictions for these low-metallicity galaxies. Equation \ref{eqone} gives very shallow high-end IMF slopes for very metal-poor galaxies, thus much depends on the exact choice of the low-metallicity cut-off. Also, for shallow high-end IMF slopes, the results become quite sensitive to the precise high-mass cut-off that is used. For an IMF defined in between 0.1 $\rm{M_{\odot}}$ and 100 $\rm{M_{\odot}}$, half of the initial mass is created in stars heavier than 50 $\rm{M_{\odot}}$ for a high-end IMF slope of zero. In equation \ref{eqone} this happens at $log_{10}(Z/{\rm Z_{\odot}})=-0.71$. At our adopted low-metallicity cut-off of $log_{10}(Z/{\rm Z_{\odot}})=-0.6$, 37\% of the stellar mass is created in stars heavier than 50 $\rm{M_{\odot}}$. It is good to keep in mind that the IMF measurements of \citet{MartinNavarro15} are still marginally consistent with a minimal high-end slope of 1.3 (Kroupa) for galaxies with $log_{10}(Z/{\rm Z_{\odot}})<-0.29$, in which case there would be a bottom-heavy IMF for metal-rich galaxies, but no top-heavy IMF for metal-poor galaxies. 

\begin{figure}
\center
\includegraphics[width=1.0\columnwidth]{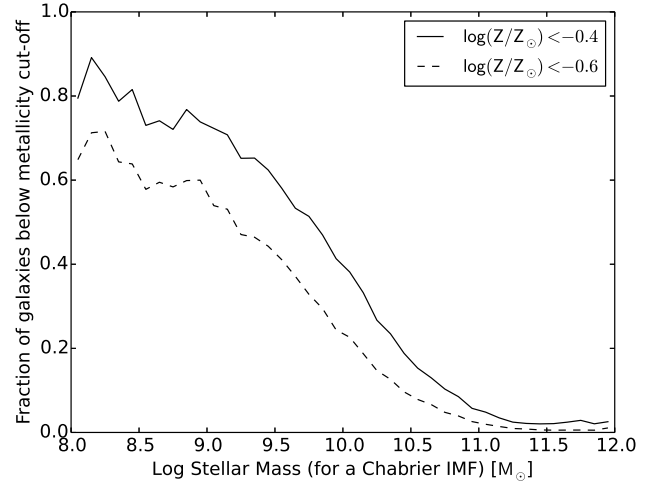}
\caption{The dashed line shows the fraction of galaxies below our adopted low-metallicity cut-off of the IMF-metallicity relation, equation \ref{eqone}, at $log_{10}(Z/{\rm Z_{\odot}})=-0.6$. For masses below $\rm{10^{9.2} M_{\odot}}$, more than 50\% of the galaxies have a metallicity below this cut-off. The solid line shows the same fraction for the lowest metallicity in the sample of \protect{\citet{MartinNavarro15}}.}
\label{FigureFractions}
\end{figure}

We may expect a top-heavy IMF for extremely low metallicities in theorised Population III stars, but the IMF in the intermediate metallicity regime is not well constrained. This metallicity regime is however probed by the IMF determination of \citet{Marks12}. They report an IMF dependence on metallicity for globular clusters that is qualitatively similar to \citet{MartinNavarro15}. Their fit of the metallicity-dependent high-end IMF slope reaches 0.05 for the lowest metallicities ($Z\approx-2.5 {\rm Z_{\odot}}$), so even lower than our adopted minimum value of 0.34. Figure \ref{FigureFractions} shows the fraction of galaxies in our sample that fall below our adopted cut-off of the IMF-metallicity relation as a function of stellar mass for a Chabrier IMF. Below a mass of $\rm{10^{9.2} M_{\odot}}$ (from SED fitting with a Chabrier IMF) 50\% of the galaxies have a metallicity lower than the cut-off. A simple extrapolation of equation \ref{eqone} would therefore enhance the effect of a top-heavy IMF at low metallicities with respect to the results reported in this work for low masses.

\subsection{The matched low-end IMF}

Another implicit extrapolation of the IMF determination from spectral indices by \citet{MartinNavarro15} is contained in the parametrisation used for the IMF. Although the free parameter of the \citet{Vazdekis96} IMF parametrisation is the high-end IMF slope, this high-end slope is not directly measured. \citet{MartinNavarro15} state that the IMF sensitive features more closely trace the dwarf-to-giant ratio $F_{0.5}$, as defined in \citet{Barbera13}: 
\begin{equation}
F_{0.5}  = \frac{\int_{0.1 \rm{M_{\odot}}}^{0.5 \rm{M_{\odot}}}M \Phi(M) dM}{\int_{0.1 \rm{M_{\odot}}}^{100 \rm{M_{\odot}}}M \Phi(M) dM} ,
\label{eqfhalf}
\end{equation}
which denotes the initial mass fraction in stars below 0.5 $\rm{M_{\odot}}$. In fact, since most ETGs contain very few stars more massive than 1.0 $\rm{M_{\odot}}$, beyond this mass the inferred IMF is more sensitive to the parametrisation than to the direct observations. It would be interesting to see whether the use of an IMF parametrisation for which the low-mass end varies in the analysis of \cite{MartinNavarro15}, would indeed give similar results in terms of the trend between metallicity and $F_{0.5}$.

In this work, apart from the Vazdekis IMF parametrisation, we will also use an alternative `low-end' IMF parametrisation in which the IMF slope below $\rm{0.5 M_{\odot}}$ is varied and the slope above  $\rm{0.5 M_{\odot}}$ is fixed at the standard Kroupa value of 1.3. Instead of matching the Vazdekis IMF slope to the low-end slope that yields the same $F_{0.5}$, we match the two IMF parametrisations by requiring the same value of $F_{0.5}/F_{1.0}$:
\begin{equation}
F_{0.5}/F_{1.0}  = \frac{\int_{0.1 \rm{M_{\odot}}}^{0.5 \rm{M_{\odot}}}M \Phi(M) dM}{\int_{0.1 \rm{M_{\odot}}}^{1.0 \rm{M_{\odot}}}M \Phi(M) dM}  \,.
\label{eqfhalfone}
\end{equation}
Matching directly on $F_{0.5}$ would not eliminate the implicit extrapolation of the Vazdekis IMF to high masses.

For a shallow Vazdekis IMF slope, $F_{0.5}$ can become very small, but this does not mean that the stars below 0.5 $\rm{M_{\odot}}$ do not give a significant contribution to the \emph{current} mass- or light-budget of the galaxy, since all the high-mass stars have turned into stellar remnants. Matching the two IMF parametrisations on $F_{0.5}/F_{1.0}$ more closely preserves the actually measured signal related to the dwarf-to-giant ratio. In principle one could choose to tune the mass parameter in $F_{M}$ separately for each galaxy, depending on its age, but we chose to fix this value at 1.0 $\rm{M_{\odot}}$. The fact that the TiO$\rm{_{2}}$ and TiO$_1$ indices in \citet{Barbera13} do not depend much on the SSP age below 10 Gyr (see their figure 4) shows that even for ETGs that have some stars with masses above 1.0 $\rm{M_{\odot}}$, these will not dominate the measured IMF signal.

Throughout this paper we will present results for the two IMF parametrisations described above, but one should keep in mind that any normalised linear combination of the two, distributing the metallicity-dependent dwarf-to-giant ratio variations over the low- and high-end IMF slopes, would also be in agreement with the data from \citet{MartinNavarro15}. Figure \ref{FigureParametrisation} shows some examples of IMFs for the two different parametrisations that are matched on $F_{0.5}/F_{1.0}$.

\subsection{Star formation rates}

In order to estimate the changes to the inferred star formation rates due to a variable IMF, we compare the magnitudes in several ultraviolet bands for a 100 Myr old constant star formation rate population, implemented with the help of the FSPS code. We do this for each galaxy in the sample, interpolating the metallicity between the discrete BaSTI values and the IMF slope on a 0.1 width grid. As SFR tracers we use the SDSS u-band, the GALEX FUV band and the total intensity of ionising radiation obtained from integrating the FSPS spectra up to 912.0 \AA\footnote{Using the number of ionising photons instead of the total energy in ionising photons gives indistinguishable results.}. These three different bands are sensitive to star formation on different time scales. The FUV-band magnitude of a galaxy with a constant star forming rate levels off after about 100 Myr (although this is an IMF dependent statement), with H-alpha (which tracks the ionising luminosity) being sensitive to shorter time scales and the u-band to much longer time scales. Part of the high energy photons that are emitted by the young massive stars will be absorbed by dust and re-emitted in the infrared. For our purposes this is a second order effect. Assuming that the infrared output from dust is in equilibrium with the input in energetic radiation from massive young stars, it scales in a similar way with IMF changes as the UV emission.

\begin{figure}
\center
\includegraphics[width=1.0\columnwidth]{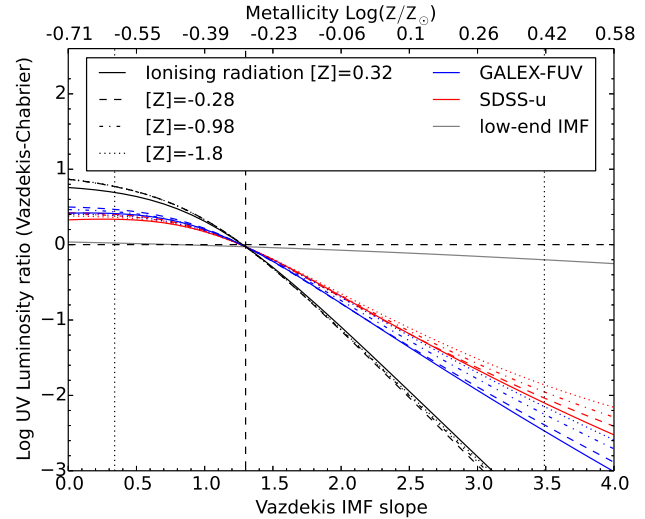}
\caption{The luminosity ratio of a 100 Myr old constant star formation rate stellar population with a Vazdekis IMF with the given high-end slope and that of the same stellar population with a Chabrier IMF. The top horizontal axis gives the metallicities corresponding to the IMF slopes via equation \ref{eqone}. Different colours depict different bands: black for for the total luminosity in ionising radiation, blue for the GALEX-FUV band and red for the SDSS-u band. Different line styles depict different metallicities. The black dashed horizontal and vertical lines depict the Chabrier values and the black dotted vertical lines show the minimum and maximum values of the Vazdekis IMF slope that are used in this work. The implied correction to the star formation rate when switching from the Chabrier to the Vazdekis IMF is the inverse of this luminosity ratio. These implied SFR changes span orders of magnitude. The much smaller luminosity ratio between the low-end IMF and the Chabrier IMF is indicated by the solid grey curve. It does not depend on the choice of UV-band.}
\label{FigureSFR}
\end{figure}

Figure \ref{FigureSFR} shows how the luminosity in different bands of a 100 Myr old constant star formation rate population changes with respect to a Chabrier IMF as a function of the assumed Vazdekis slope. There is a slight dependence on the metallicity of the stellar population (the metallicity-IMF relation has not yet been implemented here). The SDSS-u, GALEX-FUV and total ionising luminosity show similar offsets. For a shallow Vazdekis IMF slope (i.e. a low metallicity) most star formation goes into massive stars and the different tracers can indicate, if interpreted using a Chabrier IMF, a SFR up to an order of magnitude too high. For a steep Vazdekis IMF slope (i.e. a high metallicity) mostly dwarf stars are formed and the luminosity in the given bands drops dramatically by up to three orders of magnitude. The difference between the FUV luminosity and the luminosity in ionising radiation also implies a calibration offset between the FUV-SFR and H$\rm{\alpha}$-SFR that can exceed an order of magnitude\footnote{A similar offset in the H$\rm{\alpha}$-UV flux ratio appears in the IGIMF theory for low-SFR galaxies \citep{Pflamm09}.}.

For our sample of observed galaxies we adjust the SFR on a galaxy-by-galaxy basis by the inverse of the luminosity ratio of Figure \ref{FigureSFR}. For each galaxy we use the Vazdekis IMF slope implied by the IMF-metallicity relation, equation \ref{eqone}. As a conservative estimate of the implied SFR change, we use the ratio of luminosities in the GALEX-FUV band.

For the `low-end' IMF parametrisation, shown in grey in Figure \ref{FigureSFR}, the SFR changes are much smaller than for the `Vazdekis' case. Moreover, the implied changes are in this case nearly identical for the different bands and metallicities. This is because the `low-end' IMF changes the reservoir of stars that are formed below 0.5 $\rm{M_{\odot}}$, while keeping the IMF slope above 0.5 $\rm{M_{\odot}}$ fixed. Because dwarf stars with $M<0.5{\rm M_{\odot}}$ give a negligible contribution to the luminosity in these bands, the only effect is that of a change in the overall normalization of the high-mass end of the IMF.

\subsection{Stellar masses}

\begin{figure}
\center
\includegraphics[width=1.0\columnwidth]{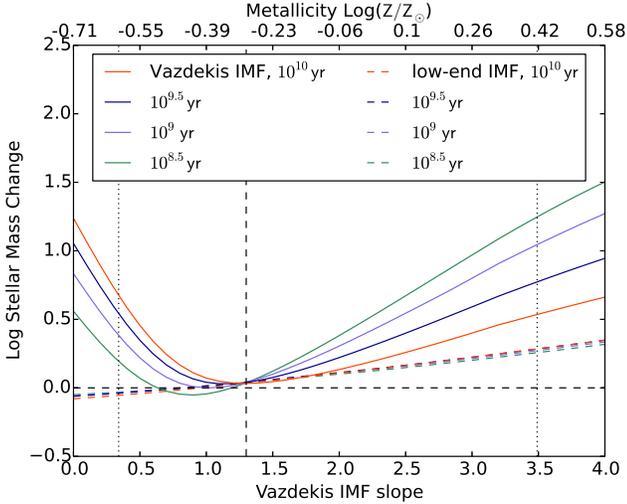}
\caption{The implied stellar mass changes of galaxies resulting from switching from a Chabrier IMF to a metallicity-dependent IMF, equation \ref{eqone}, for a fixed observed luminosity. The double horizontal axis reflects this equation. The mass changes are based on the change in mass-to-light ratio in the Cousins R-band for a galaxy with a fixed age and metallicity. Different line colours represent different SSP ages. The dashed lines represent the low-end IMF parametrisations matched to the Vazdekis IMF on $F_{0.5}/F_{1.0}$, equation \ref{eqfhalfone}. The black dashed lines depict the Chabrier values and the black dotted vertical lines show the minimum and maximum values of the Vazdekis IMF slope used in this work. The stellar mass increase at high metallicities is due to dwarf stars. The slight stellar mass decrease for low metallicity  and a low-end IMF is due to a shortage of dwarf stars. The Vazdekis IMF shows an increase in mass in this region due to stellar remnants. A similar plot for a constant star formation rate instead of a simple stellar population is shown in the Appendix, Figure \ref{FigureMassChangeB}.}
\label{FigureMassChange}
\end{figure}

Figure \ref{FigureMassChange} shows the galaxy mass changes caused by a switch from a Chabrier IMF, depending on metallicity and r-band weighted age. The mass change is determined by the change in mass-to-light ratio in the Cousins R-band. Galaxies are represented by a simple stellar population. This is a much better approximation for the ETGs used in the IMF determination of \citet{MartinNavarro15} than for actively star forming galaxies. This simplification of the star formation history is justified in the Appendix by comparing results from a SSP with those from a constant star formation history. Qualitative differences only show up for low-metallicity galaxies in the Vazdekis IMF parametrisation. 

Figure \ref{FigureMassChange} shows that for steep Vazdekis IMF slopes, or equivalently high metallicities, switching from a Chabrier IMF to a Vazdekis IMF can increase the inferred stellar mass by more than an order of magnitude due to an increase in the fraction of dwarf stars. Most of these metal-rich galaxies are old though, which results in a mass change of approximately 0.6 dex. For metal-poor galaxies, the increased fraction of stellar remnants can induce a mass increase of $\rm{0.2-0.65}$ dex. For the low-end IMF the expected stellar mass changes are much smaller, ranging from -0.06 dex at low metallicities due to a shortage of dwarf stars to 0.29 dex at high metallicities due to an excess of dwarf stars. Mass changes for the low-end IMF are almost independent of age.

It is worth noting that for old galaxies a high-end IMF slope around 1.3, or equivalently a Chabrier-like IMF, minimizes the inferred stellar mass of the galaxy. This means that any variations in the high-end slope of the IMF away from Chabrier are bound to increase the inferred mass of the galaxy. A Chabrier IMF (or a similar Kroupa/Canonical IMF) balances the amount of mass locked up in dwarf stars on the one side and stellar remnants on the other side in such a way that the total of this invisible stellar mass is close to minimal at late times.

\section{Star Formation Main Sequence}
\label{SectionMainSequence}

\begin{figure*}
\center
\includegraphics[height=0.90\columnwidth]{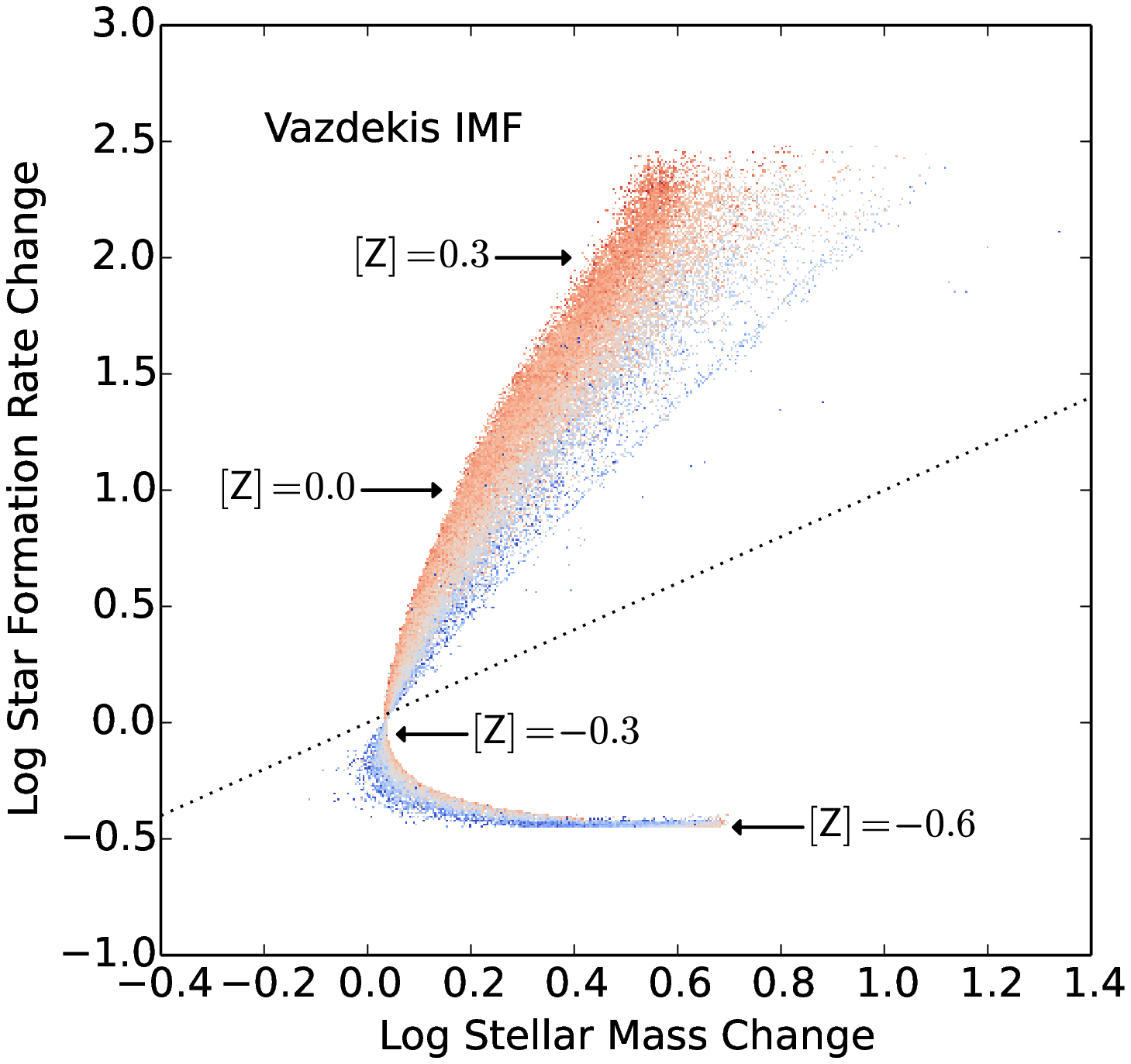}
\includegraphics[height=0.90\columnwidth]{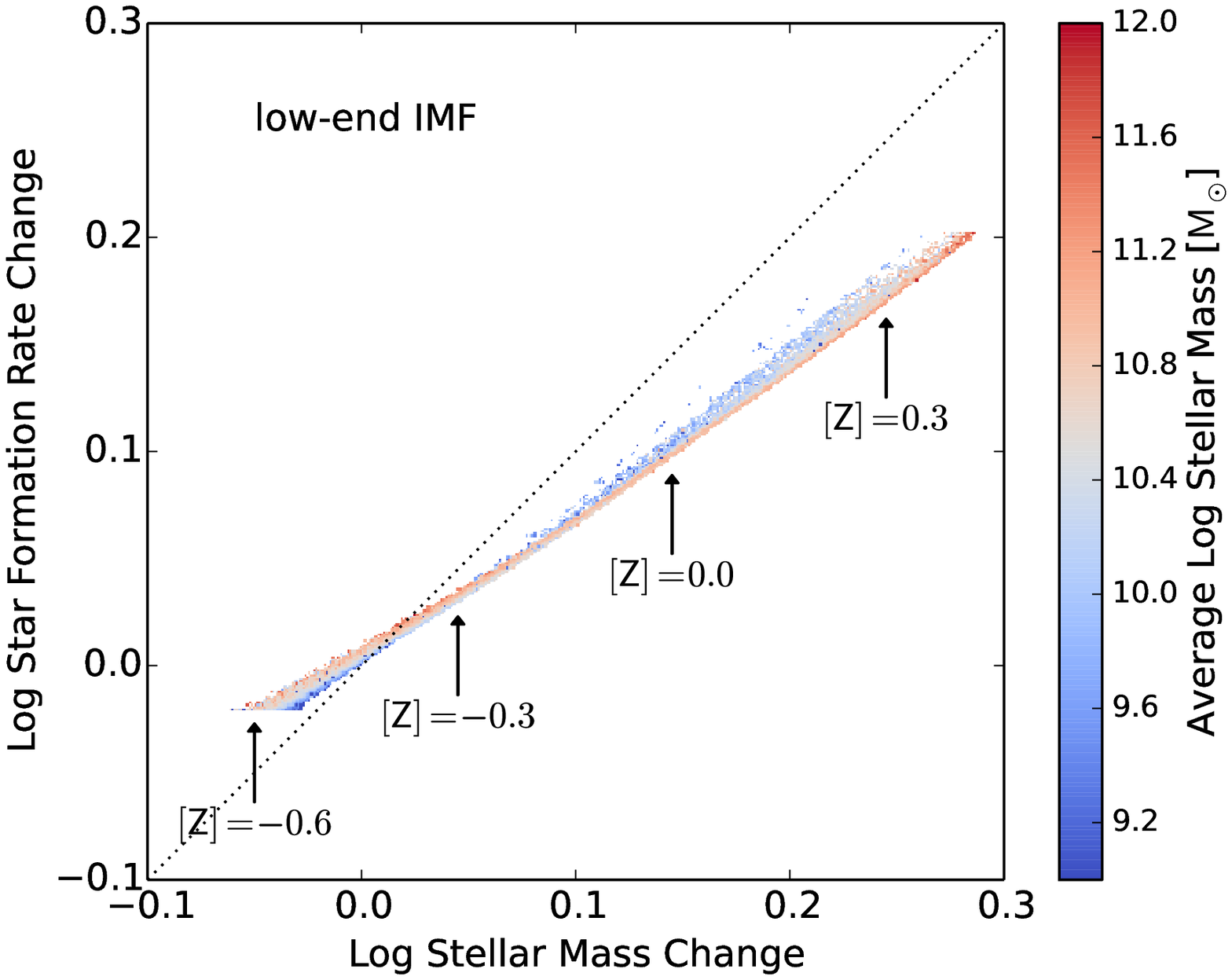}
\caption{The implied stellar mass change and star formation rate change with respect to observations that assume a Chabrier IMF for a sample of SDSS galaxies that obey the IMF-metallicity relation of \protect{\citet{MartinNavarro15}}, equation \ref{eqone}, with metallicities from \protect{\citet{Gallazzi05}}. The colour scale indicates the average log stellar mass of the galaxies in each pixel. Some typical metallicity values are indicated with arrows. The black dotted line depicts a constant specific star formation rate. The left panel assumes the Vazdekis IMF parametrisation for which the free parameter is the high-end IMF slope. The right panel assumes our low-end IMF parametrisation, for which the free parameter is the low-end IMF slope. Both panels show large changes, but the `Vazdekis' case has much more drastic consequences, with star formation rate changes of over two orders of magnitude. At a given metallicity, any normalised linear combination of these two IMFs would be in agreement with the IMF measurements from line indices by \protect{\citet{MartinNavarro15}}. }
\label{FigureMassDifSfrDif}
\end{figure*}

Figure \ref{FigureMassDifSfrDif} shows the mass and SFR changes relative to a Chabrier IMF implied by the metallicity-dependent IMF from \citet{MartinNavarro15}, equation \ref{eqone}, for 186,886 SDSS galaxies. The left panel shows the results for the Vazdekis IMF parametrisation and the right panel shows the results for the low-end IMF parametrisation. The large difference between both panels clearly shows that the effects of the IMF change is more sensitive to the IMF parametrisation, rather than to the direct measurements of \citet{MartinNavarro15}.

The `Vazdekis IMF' panel shows two tails emanating from the (zero SFR change, zero mass change) point. The lower tail consists of low-metallicity galaxies, which have a shallow high-end IMF slope. This results in a negative SFR correction, since there are fewer dwarf stars formed than for a Chabrier IMF. The stellar mass correction for most of these galaxies is positive though, because of invisible mass locked up in stellar remnants. The maximum extent of this effect depends in part on the assumed cut-off in equation \ref{eqone} below $log_{10}(Z/{\rm Z_{\odot}})=-0.6$. The upper tail consists of high-metallicity galaxies, which have a steep high-end IMF slope. These have more star formation in dwarf stars than for a Chabrier IMF, which results in an increase of the inferred star formation rate of up to two orders of magnitude. The increase in stellar mass is smaller, which means that the specific star formation rate (SSFR) is increased as well.

The effect for the `low-end' IMF parametrisation is much smaller (especially for the SFR) but still significant. In this case the whole effect can be understood in terms of how much mass is locked up in dwarf stars, since the IMF slope above 0.5 $\rm{M_{\odot}}$ is kept fixed. Naively one might think that the specific star formation rate should remain unchanged in this case, since the percentage of extra dwarf stars formed in current star formation is the same as the percentage of extra dwarf stars locked up from older generations. This would be true if the SSFR were defined as the SFR divided by the total initial stellar mass formed, but it is instead defined with respect to the current stellar mass \citep{Hopkins08}. The current total stellar mass is lower than the initial stellar mass, because of mass loss by intermediate-mass and massive stars. This means that for an IMF with relatively more dwarf stars, there is an extra relative `increase' of stellar mass with respect to the Chabrier case, caused by a diminished stellar mass loss (and thus a diminished difference between the initial stellar mass and the current stellar mass). This means that the change in stellar mass is larger than the change in SFR and thus that the SSFR decreases. The opposite happens in the case where the `low-end' IMF has fewer dwarf stars than the Chabrier IMF. This causes the pattern in the second panel of Figure \ref{FigureMassDifSfrDif}, which shows the displacement of the galaxies in the mass-SFR plane. The change in SFR is roughly two thirds of the change in mass.

\begin{figure*}
\center
\includegraphics[height=0.65\columnwidth]{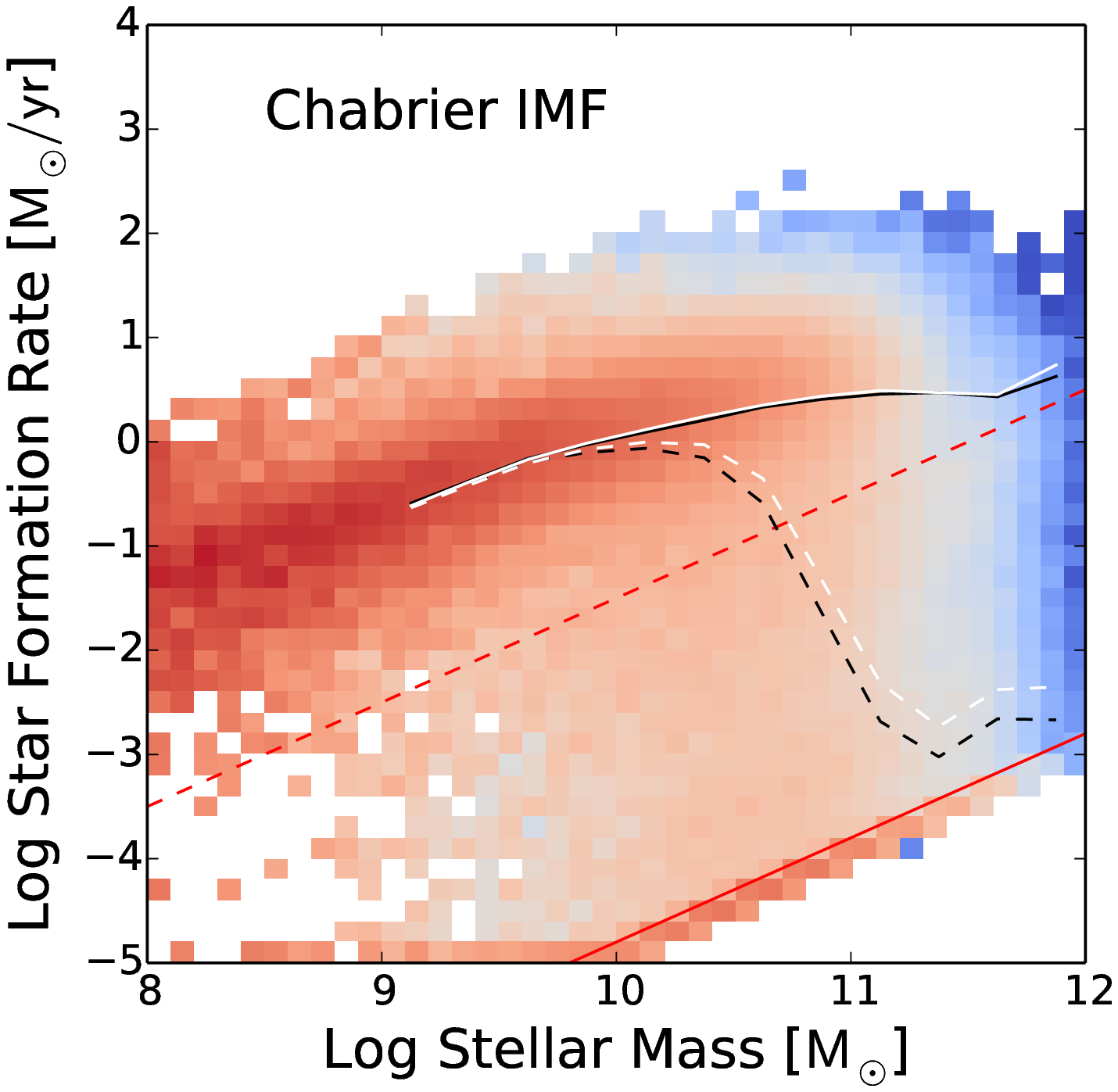}
\includegraphics[height=0.65\columnwidth]{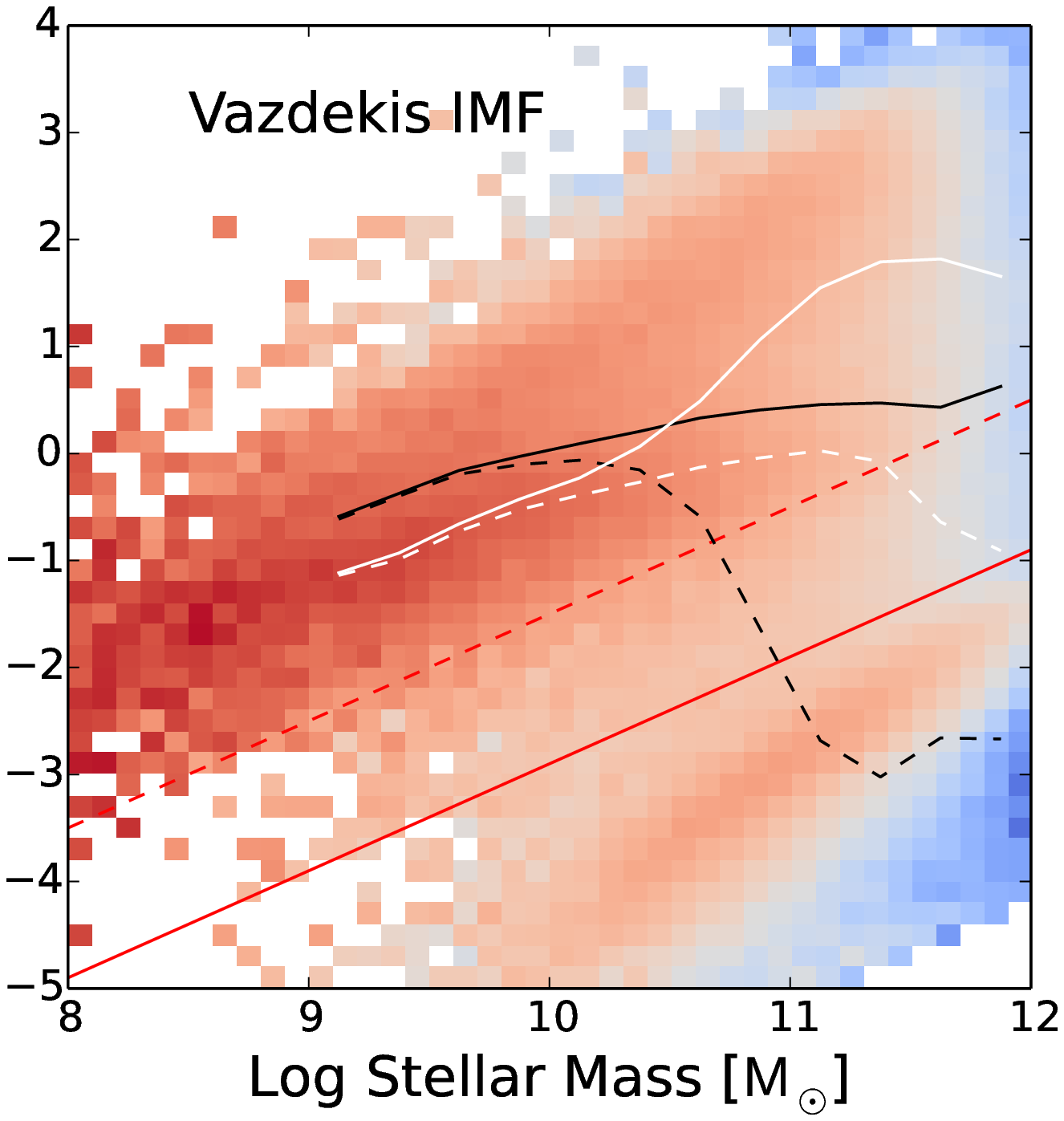}
\includegraphics[height=0.65\columnwidth]{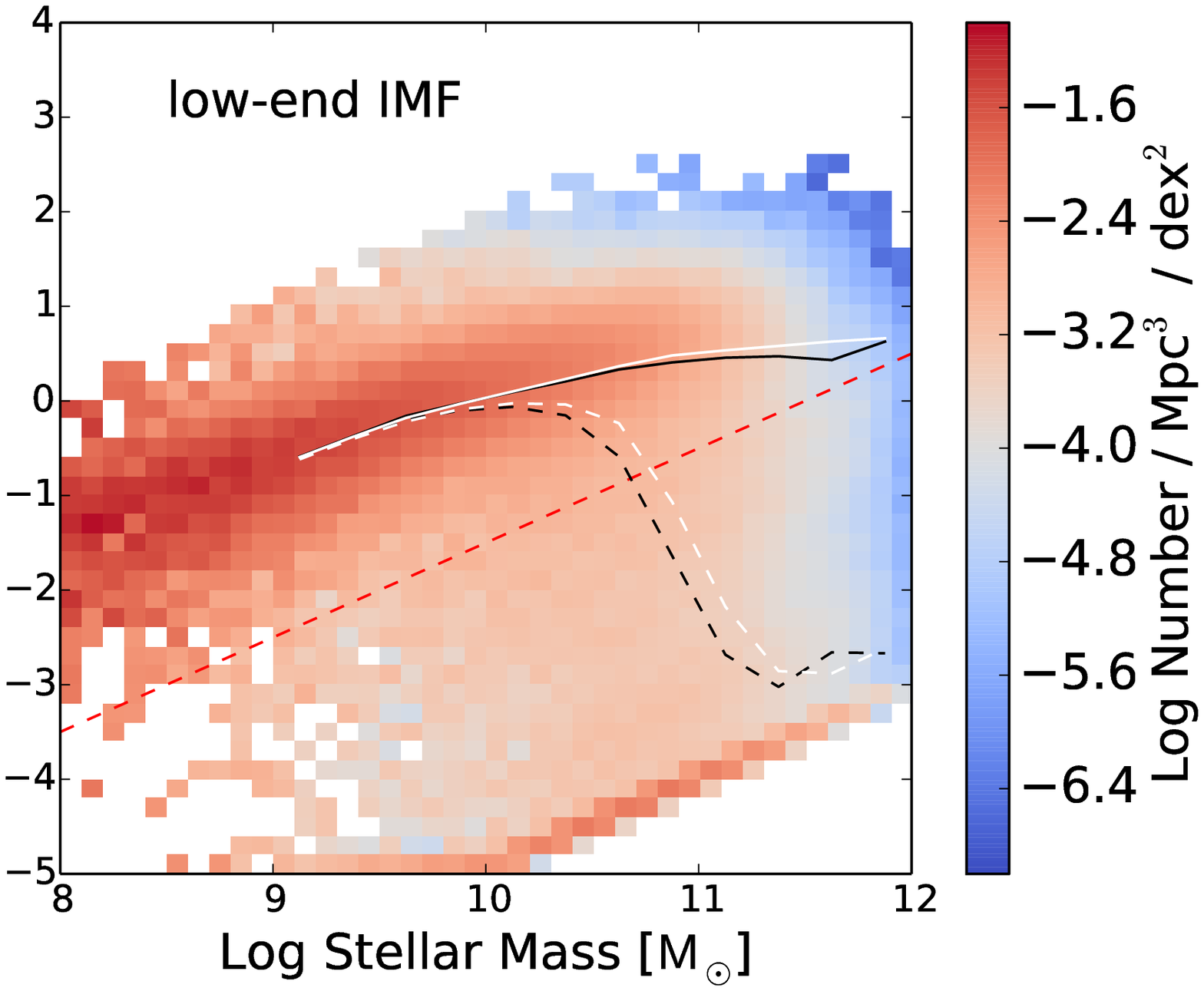}
\caption{The inferred star formation rate as a function of inferred stellar mass for different IMFs. The left panel shows the matched sample of SDSS galaxies with metallicities from \protect{\citet{Gallazzi05}} and masses and star formation rates from \protect{\citet{Chang15}}, assuming a Chabrier IMF. The middle panel shows the results for the metallicity-dependent Vazdekis IMF, as described in Figure \ref{FigureMassDifSfrDif}. The right panel shows the effect of a low-end metallicity-dependent IMF. For ease of comparison, the medians of the matched sample with the Chabrier IMF have been plotted in black in all panels: dashed lines denote the median for all galaxies and solid lines denote the median for all star-forming galaxies (defined as those galaxies with a SSFR larger than $\rm{10^{-11.5}/yr}$, indicated with a red dashed line). White lines in the left panel show the medians for the complete sample from \protect{\citet{Chang15}}. In the middle and right panels the white lines show the medians of the underlying distributions. The excellent agreement between the white and black solid lines in the left panel shows that our selection on galaxies for which stellar metallicities are available, does not introduce a bias. The red solid line in the left panel demarcates the boundary below which the star formation rates should be regarded as upper bounds. The region below the red solid line in the middle panel is contaminated by these upper bounds. The middle panel shows that the metallicity-dependent Vazdekis IMF has a large influence on the mass-SFR relation, while the low-end metallicity-dependent IMF in the right panel has only a small effect. Any normalised linear combination of these two cases would be in agreement with the IMF measurements from line indices by \protect{\citet{MartinNavarro15}}.}
\label{FigureMassSfr}
\end{figure*}

Figure \ref{FigureMassSfr} shows the effect of the metallicity-dependent IMF on the star formation main sequence. The left panel shows the mass-SFR relation for the selection of 186,886 galaxies out of the 858,365 SDSS galaxies of \citet{Chang15} that have a well-defined metallicity and age determination by \citet{Gallazzi05}. The running median SFR of star-forming galaxies with $SSFR > 10^{-11.5} /{\rm yr}$, black solid line, coincides with that for the complete \citet{Chang15} sample (solid white line). Therefore, we expect no significant selection effects for the main sequence. The running median for the complete sample (black and white dashed lines) shows a slight underrepresentation of quiescent galaxies in our sample with respect to the complete \citet{Chang15} sample. The colour scale in Figure \ref{FigureMassSfr} represents the number of galaxies per $\rm{Mpc^3}$, per dex stellar mass, per dex SFR. Individual galaxies have been weighted by $1/V_{max}$ from \citet{Chang15}, correcting for the maximum volume in which the given galaxy would be selected. The whole sample has been rescaled by 858,365/186,886, neglecting the aforementioned slight underrepresentation of quiescent galaxies in our sample.

The middle and right panels show the effect of the metallicity-dependent dwarf-to-giant ratio for respectively a Vazdekis IMF parametrisation and a low-end IMF parametrisation, as described in section \ref{SectionMethod}. The Vazdekis IMF has a dramatic influence on the star formation main sequence (compare the black and white solid lines in the middle panel). Below a mass of $\rm{\sim 10^{10.5} M_{\odot}}$ the dominant effect is from metal-poor galaxies which lower the median inferred SFR by $\rm{\sim 0.5}$ dex relative to a Chabrier IMF. Above this mass the dominant effect is from metal-rich galaxies which increase the median SFR, effectively extending the SSFR main sequence up to a stellar mass of $\rm{\sim 10^{12} M_{\odot}}$ and making it more linear. We used a running median to represent the typical galaxy at each mass. The true average SFR at low masses is actually increased, because of an increase in the scatter.

The increased scatter in this plot could indicate that the underlying metallicity-IMF relation or the Vazdekis parametrisation of the IMF is wrong, because in general a more precise description will cause a decrease in the observed scatter. This is not necessarily the case though. Apart from measurement or modelling errors in the metallicity, the increased scatter could also be an intrinsic property of the galaxy population. For a galaxy population that truly follows the assumed Vazdekis IMF-metallicity trend, the middle panel would indicate the total star formation, but the left panel would still indicate the effective high-mass star formation. It is possible that the high-mass star formation is better behaved than the total star formation. For a varying IMF these two quantities are no longer directly linked to each other, but it is the high-mass star formation that determines the stellar feedback. It could well be that this causes the high-mass star formation to be regulated, whereas the increased dwarf star formation for metal-rich galaxies could create a runaway process of star formation that is less regulated and therefore shows a larger galaxy-by-galaxy variation at a given mass. We will discuss this possibility in section \ref{SectionQuenching}.

The right panel shows results for the low-end IMF parametrisation. For this parametrisation there is almost no effect at low masses and at high masses the star formation main-sequence is prolonged to slightly higher masses. The effect on the median for star forming galaxies is especially small, since individual galaxies are displaced almost parallel to lines of constant SSFR, along the main sequence, see also Figure \ref{FigureMassDifSfrDif}. In practice one could construct an IMF parametrisation in between the two options described here, for which part of the observed trend of the IMF sensitive line indices would be caused by a changing low-end IMF slope and part would be caused by a changing high-end slope. The effects on the mass-SFR plot would fall somewhere in between the middle and right panels of Figure \ref{FigureMassSfr}. Qualitatively this means an extension of the star formation main sequence together with a possible lowering of the median SFR at low masses, but quantitatively the magnitude of this effect is not well constrained by current observations.

\begin{figure}
\center
\includegraphics[width=1.0\columnwidth]{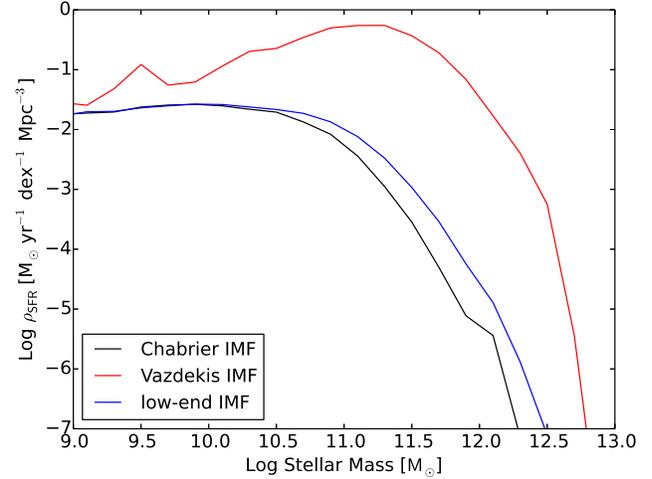}
\caption{The inferred star formation rate density as a function of the inferred stellar mass for the sample of SDSS galaxies with different IMFs. The black line shows the original estimate for a Chabrier IMF. The red line corresponds to the metallicity-dependent Vazdekis IMF and the blue line corresponds to the metallicity-dependent low-end IMF. The increased number of dwarf stars for metal-rich galaxies causes a very large increase in the inferred SFR, which peaks around a mass of $\rm{10^{11.3} M_{\odot}}$ for the Vazdekis IMF. For the low-end IMF the effect is much more subtle.}
\label{FigureSfrDensity}
\end{figure}

Figure \ref{FigureSfrDensity} shows the inferred low-redshift SFR densities of the Universe as a function of galaxy mass. The effect of a low-end metallicity-dependent IMF is quite small, shifting the star formation to higher masses by about 0.2 dex. The implied increase of the total SFR density of galaxies more massive than $\rm{10^{9} M_{\odot}}$ is a factor of 1.11. For the Vazdekis IMF this is a factor of 16.6. The effect is much stronger, with a large peak of dwarf star formation around $M_{*,SSP}\approx10^{11.3} {\rm M_{\odot}}$.

\section{Galaxy Stellar Mass Function}
\label{SectionGSMF}

\begin{figure}
\center
\includegraphics[width=1.0\columnwidth]{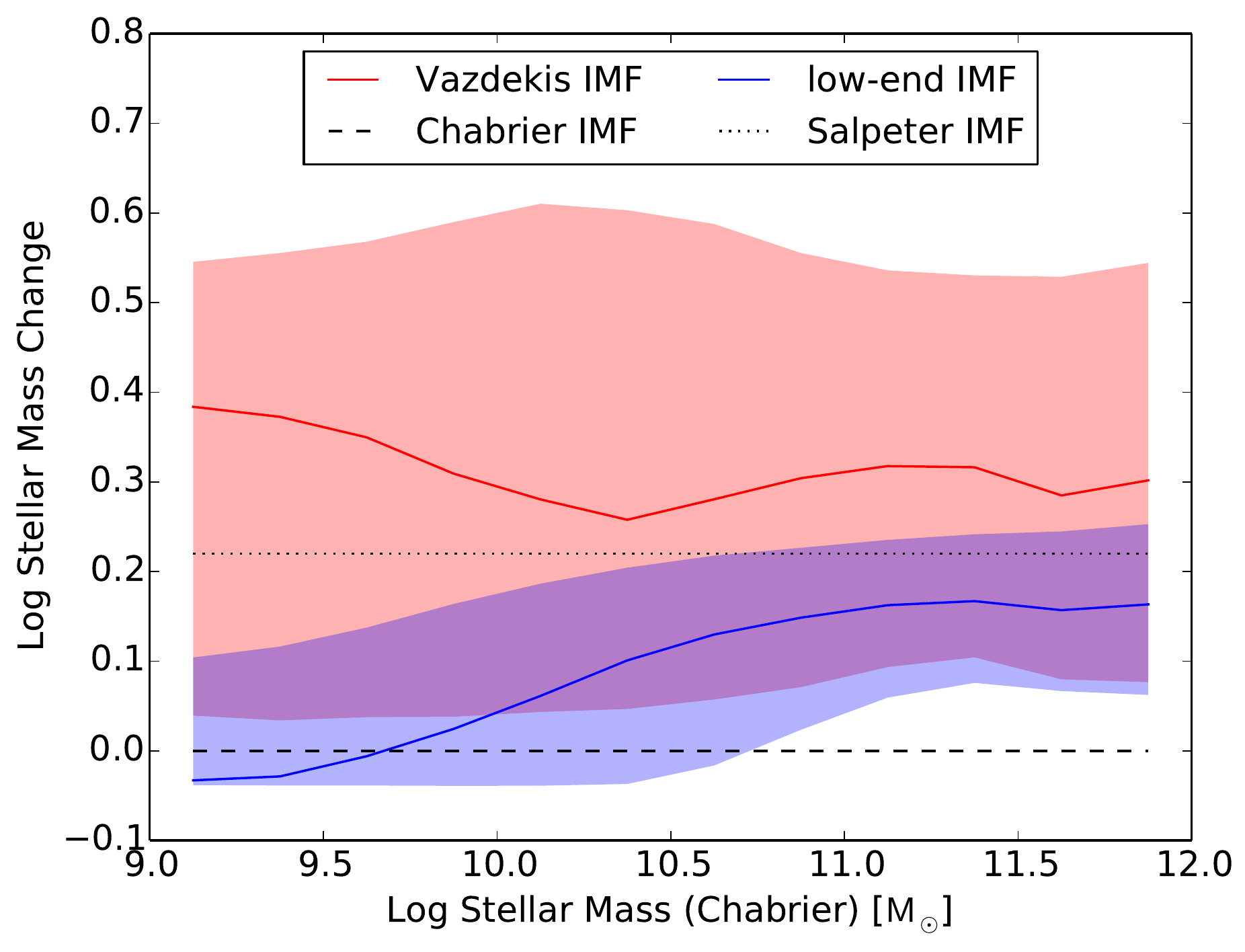}
\caption{The difference in stellar mass estimates between a metallicity-dependent IMF and a Chabrier IMF as a function of the stellar mass inferred for a Chabrier IMF for the sample of 186,886 SDSS galaxies. The red line and shaded region show the median and the 10\%-90\% region for the Vazdekis IMF. Blue corresponds to the low-end IMF. The black dashed and dotted lines depict the Chabrier and Salpeter IMFs respectively.}
\label{FigureMassShifts}
\end{figure}

Figure \ref{FigureMassShifts} shows the mass change due to the change from a Chabrier IMF to a metallicity-dependent IMF for our sample of galaxies as a function of  the original Chabrier stellar mass. The median mass increase for the Vazdekis IMF is larger than that for a Salpeter IMF, which might be in tension with some of the mass measurements from strong lensing and from dynamics. This suggests that at least a fraction of the IMF-metallicity trend measured by \citet{MartinNavarro15} should be provided for by a change in the low-end slope of the IMF. For the Vazdekis IMF the mass change at low masses is due to stellar remnants while at high masses it is due to dwarf stars. Nevertheless, the overall effect is insensitive to mass, with a large scatter at all masses. Thus, metallicity-dependent IMF changes do not necessarily create a clear trend of $M_{*true}/M_{*Chabrier}$ with $M_{*Chabrier}$, where the latter is obtained through SED fitting. The low-end IMF parametrisation gives smaller mass changes, but it does produce a clear trend of $M_{*true}/M_{*Chabrier}$ with $M_{*Chabrier}$, with a progressively larger IMF-induced mass correction towards higher masses.

\begin{figure}
\center
\includegraphics[width=1.0\columnwidth]{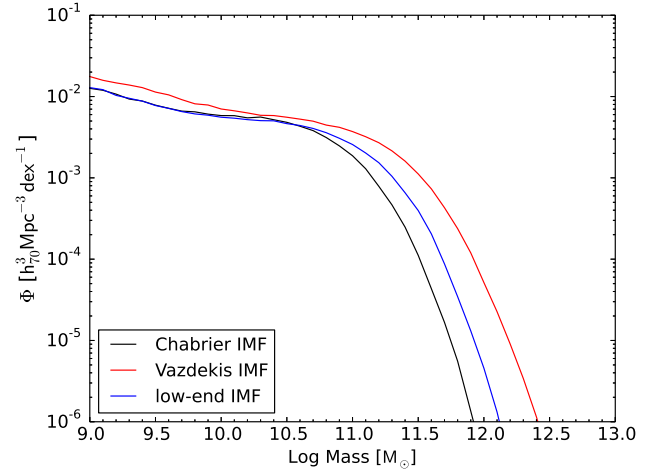}
\caption{The galaxy stellar mass function inferred from observations of \protect{\citet{Moustakas13}} for several IMFs: Chabrier (black), metallicity-dependent Vazdekis IMF (red) and metallicity-dependent low-end IMF (blue). For both metallicity-dependent IMFs the knee of the mass function shifts to higher masses, but the steepness of the drop-off is hardly affected.}
\label{FigureGSMF}
\end{figure}

In figure \ref{FigureGSMF} we show the effect on the galaxy stellar mass function inferred from the observations of \citet{Moustakas13}. Galaxies are scattered as a function of mass according to the exact distribution underlying Figure \ref{FigureMassShifts}. For galaxies below a mass of $\rm{10^{9} M_{\odot}}$ we extrapolate the Moustakas mass function with a slope of 0.41 and we apply the mass changes corresponding to the lowest 0.1 wide mass bin at $\rm{10^{9} M_{\odot}}$. For both IMF parametrisations the knee of the mass function shifts to a higher mass, but the steepness of the high-mass drop-off is not affected much. The abrupt nature of this drop-off is generally believed to be caused by the feedback of active galactic nuclei (AGN). This AGN feedback should be very effective at quenching high-mass galaxies in order to prevent more massive galaxies from forming. A metallicity-dependent IMF does not remove the need for such an efficient quenching mechanism at high masses, it only shifts the mass at which this quenching occurs, which can also be seen in Figure \ref{FigureMassSfr}. The IMF changes also imply a change in the total stellar mass density of the Universe. The total stellar mass density in galaxies above $\rm{10^{9} M_{\odot}}$ increases by a factor 1.33 for the low-end IMF and by a factor 2.31 for the Vazdekis IMF.

In this work we focus on low-redshift observations, but the implications of a metallicity-IMF trend at higher redshifts could be very significant. A more top-heavy IMF at higher redshifts, implied by lower metallicities, would affect the relation between the total star formation rate density implied by direct SFR measurements at these redshifts and that from comparing the galaxy stellar mass function at different redshifts. At high redshifts the true star formation rate would be smaller than that inferred from SFR measurements with a Chabrier IMF. The star formation rate density, deduced from comparing neighbouring redshift slices of the galaxy stellar mass function, interpreted with a Chabrier IMF, would also be too high. In the case of a `low-end' IMF, the offsets in both SFR density determinations are the same, since they are caused by the same overestimation of mass locked in invisible dwarf stars with masses smaller than $\rm{0.5 M_{\odot}}$. The consistency between both SFR density determinations is thus not affected. For a Vazdekis IMF, the direct SFR density measurement is more sensitive to the IMF change than the SFR density estimate from comparing GSMFs, because it is more sensitive to the highest mass stars. In this case we thus expect a smaller increase in stellar mass in the Universe from comparing GSMFs than that implied by direct SFR observations, if both are interpreted with a Chabrier IMF.

\section{Metals}
\label{SectionMetals}

\begin{figure*}
\center
\includegraphics[height=0.92\columnwidth]{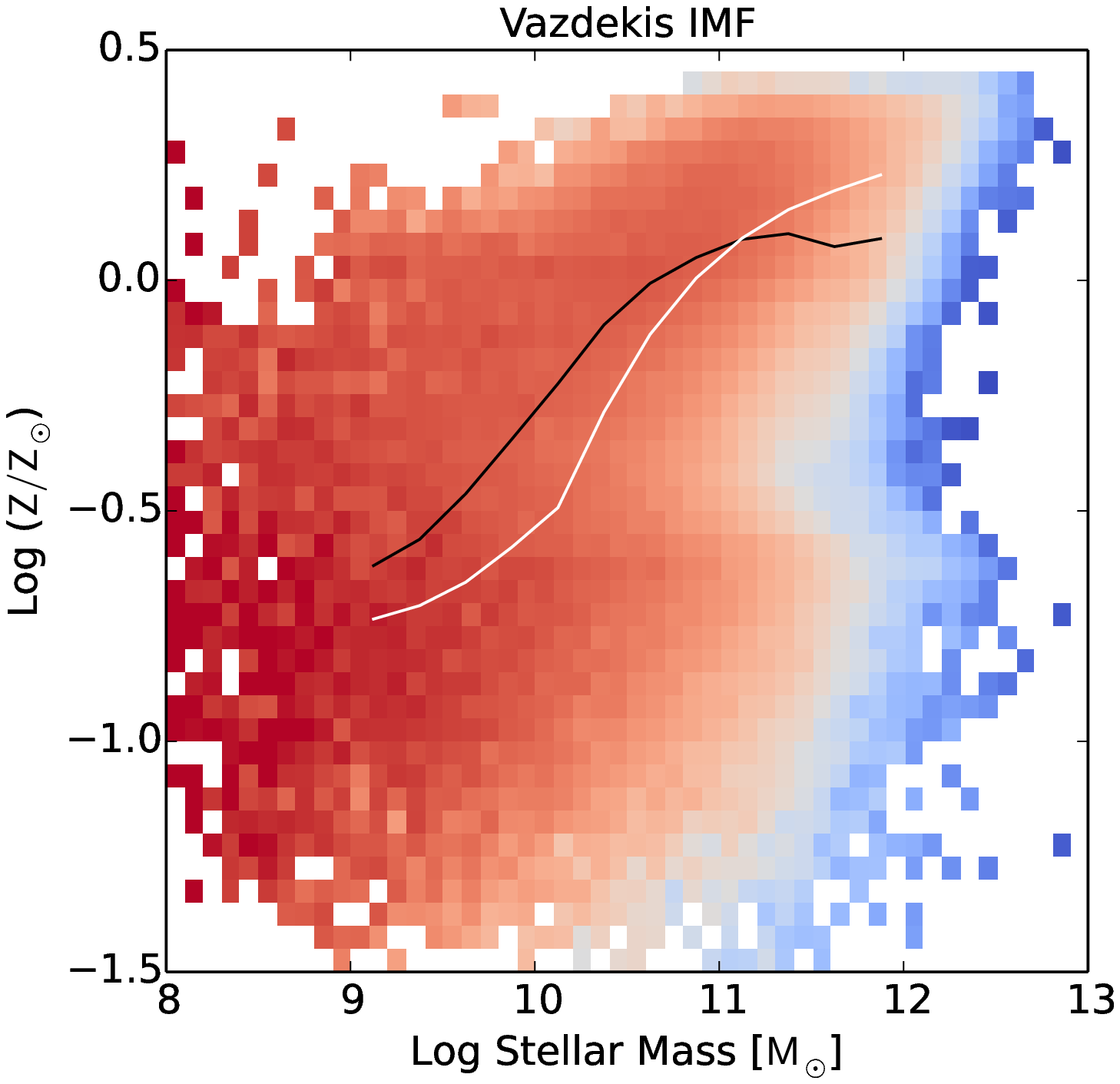}
\includegraphics[height=0.92\columnwidth]{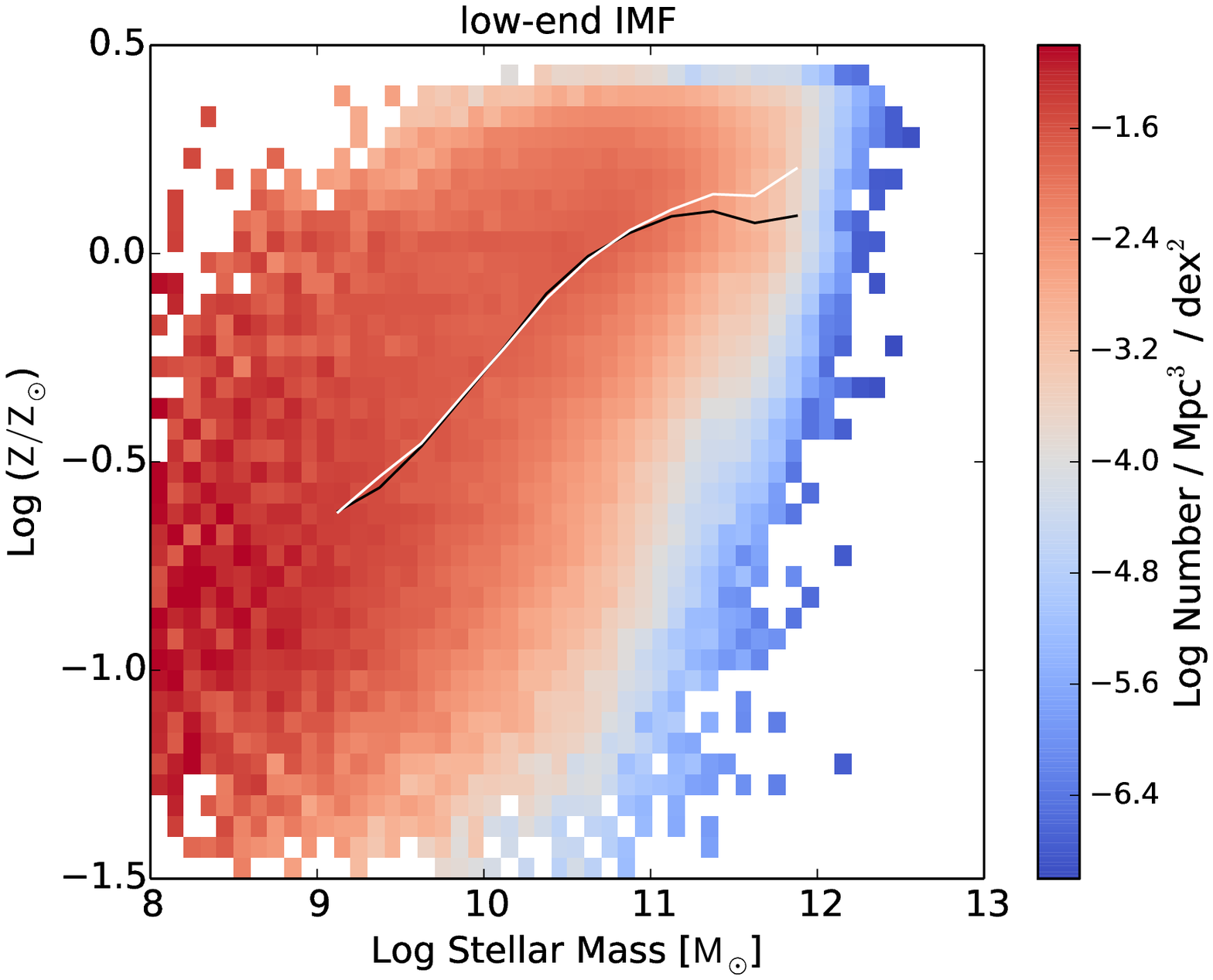}
\caption{The effect of a metallicity-dependent IMF on the inferred stellar mass-metallicity relation from \protect{\citet{Gallazzi05}}. The left and right panels show the results for respectively the Vazdekis and the low-end IMF parametrisations. White solid lines indicate the running median and the black solid line indicates the running median for a universal Chabrier IMF. In both panels the mass-metallicity relation shows considerably less flattening at high masses for the variable IMF. In the `Vazdekis IMF' panel there is a clear feature of galaxies shifting to the right at $log_{10}(Z/{\rm Z_{\odot}})\approx-0.6$. The reason that this shift remains constant below $log_{10}(Z/{\rm Z_{\odot}})=-0.6$ is that we do not extrapolate the IMF-metallicity relation, equation \ref{eqone}, below this metallicity, which is a somewhat arbitrary choice. This mass shift is due to stellar remnants that are created by a presumably more top-heavy IMF at low metallicities. The `low-end' panel does not have this feature. The effect of extrapolating below $log_{10}(Z/{\rm Z_{\odot}})=-0.6$ in this panel would be a slight shift to the left instead of to the right.}
\label{FigureMassMetallicity}
\end{figure*}

We used the stellar mass-metallicity measurements of \citet{Gallazzi05} as input for the metallicity-dependent IMF. However, a metallicity-dependent IMF will alter the stellar masses of these observed galaxies and will therefore change the mass-metallicity relation. Figure \ref{FigureMassMetallicity} shows the effect that this has for both the Vazdekis IMF and the low-end IMF. Solid white lines in both panels show the running median metallicity for the Vazdekis (left panel) and low-end (right panel) IMFs, while the black curve indicates the median metallicity for a universal Chabrier IMF. In both cases the mass-metallicity relation is steeper at high mass for the metallicity-dependent IMF than for the Chabrier IMF. We used the same weighting with $1/V_{max}$ as in Figure \ref{FigureMassSfr}.

The behaviour at low metallicity is different for the two parametrisations. In the Vazdekis case we see a clear shift to higher masses for $Z < {\rm Z_{\odot}}$. At the low-metallicity end this shift is caused by a progressively more top-heavy IMF for lower metallicities and the resulting mass gain is comprised of stellar remnants. We did not extrapolate the IMF metallicity trend, equation \ref{eqone}, below $log_{10}(Z/{\rm Z_{\odot}})=-0.6$, otherwise this shift to the right in the left panel of figure \ref{FigureMassMetallicity} would get more prominent toward lower metallicities, as can be seen in Figure \ref{FigureMassChange}.

The mass-metallicity relation inferred from the low-end IMF is much less sensitive to this extrapolation and the effect goes in the opposite direction. Since in this case the high-end IMF slope is kept fixed, the net effect of extrapolating to lower metallicities would be to have an IMF that is increasingly deficient in $M<0.5{\rm M_{\odot}}$ stars. The resulting galaxies would be less massive than suggested by SED fitting with a Chabrier IMF. This would shift the galaxies in the right panel of figure \ref{FigureMassMetallicity} at metallicities below $log_{10}(Z/{\rm Z_{\odot}})=-0.6$ slightly to the left. This effect is much less drastic than for the Vazdekis case as it is clearly bounded from below by having no stars below 0.5 $\rm{M_{\odot}}$, whereas in the Vazdekis case the ratio of observed stars over remnants is not bounded. 

\begin{figure}
\center
\includegraphics[width=1.0\columnwidth]{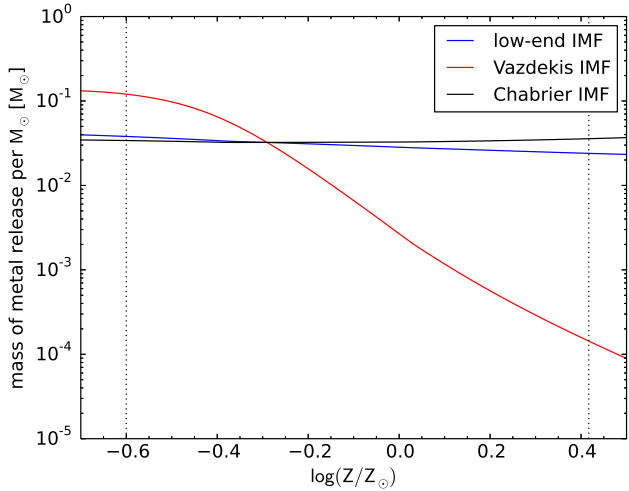}
\caption{The total metal mass released by a simple stellar population over its lifetime as a function of metallicity for the metallicity-dependent Vazdekis IMF, the corresponding low-end IMF and the Chabrier IMF.  The metal release for different stellar masses and stellar stages are taken from \protect{\citet{Portinari98}} and \protect{\citet{Marigo01}}. The trends are primarily due to changes in the IMF with metallicity and to a lesser extent due to a metallicity-dependence in the metal output of individual stars. Dotted vertical lines show the minimum and maximum values of the metallicity used in this work. }
\label{FigureYields}
\end{figure}

Figure \ref{FigureYields} shows the metal mass released over the lifetime of a simple stellar population as a function of metallicity. In order to construct this we used the metallicity-dependent yield tables for individual stars of different masses from \citet{Portinari98} and \citet{Marigo01}. There are two ways in which the metal release of a simple stellar population depends on metallicity. First, through a metallicity dependence of the metals released by individual stars, but the more significant contribution comes from the metallicity dependence of the IMF, especially for the Vazdekis IMF. In this case, a galaxy would presumably start out on the left, with a low metallicity, a top-heavy IMF and a high metal output. During the lifetime of a galaxy on average its metallicity will grow and as a result it will shift to the right in this graph, towards a more bottom-heavy IMF and a lower metal output. Apart from a boost in the metal release, the early top-heavy IMF will leave not much evidence, because these metal producing stars will have been turned into stellar remnants. The metal output inferred from the integrated contribution of SFR observations at higher redshifts over time would not change drastically though, since these SFR observations effectively measure the formation rate of high-mass stars. These are exactly the stars that produce the metals.

The same situation applies to ionising radiation, because the bulk of the ionising radiation and metals are produced by the same massive stars. In principle reionisation would get a head start due to an early top-heavy IMF. For redshifts where we have representative H$\rm{\alpha}$-based SFR measurements, the IMF-metallicity relation should not change the implied ionising photon production much, but for higher redshifts the IMF effect on reionisation can potentially be large.

\section{Rapid Galaxy Quenching}
\label{SectionQuenching}

The distribution of galaxies in a colour-magnitude plot is observed to follow a bimodal pattern \citep[e.g.][]{Strateva01,Bell04}, with most galaxies being classified as either a `red sequence' or a `blue cloud' galaxy. The red sequence is comprised of red, mostly elliptical galaxies and the blue cloud is comprised of blue, mostly spiral galaxies. Galaxies that fall in between those categories are called `green valley' galaxies. They are believed to be mostly galaxies that are currently quenching, making the transition from the blue cloud to the red sequence \citep[e.g.][]{Trayford16}.

A similar classification of galaxies can be made based on stellar mass and specific star formation rate, although this may depend on the type of SFR estimator used. A short-time SFR indicator like H$\rm{\alpha}$ may be inherently more bursty, so a quiescent galaxy based on the H$\rm{\alpha}$ SFR may not be quiescent based on the u-band SFR. Regardless of this ambiguity in defining a green valley galaxy, in general relatively few galaxies are in this `in between' state. This suggests that the quenching of a typical galaxy is a fast process. Also, especially at high masses, quenching needs to be very effective to explain the sharp drop-off in the number density of high-mass galaxies beyond the knee of the galaxy mass function.

A bottom-heavy IMF at the late stages of galaxy evolution could help speed up the last phase of the quenching process, because it produces less feedback per unit mass of stars formed and because it shortens the gas consumption time scale inferred from the observed galaxy properties. Figure \ref{FigureMassDifSfrDif} shows that a metallicity-dependent IMF can cause the inferred SFR  to increase by of up to 2.5 orders of magnitude relative to a Chabrier IMF (although the implied SFR changes are much smaller if the IMF changes apply to the low-end IMF slope rather than the high-end IMF slope). The implied gas consumption timescale\footnote{The factor of 1.33 captures the contribution from helium.}, $\tau_{gas}=1.33(M_{HI}+M_{H_2})/SFR$, i.e. the timescale for the galaxy to run out of its current cold gas reserve, scales with the inverse of the SFR and will thus be significantly shortened. On the other hand, the expected stellar feedback for a galaxy with such a bottom-heavy IMF will not change much, because the observed SFR already effectively represents the SFR in high-mass stars.

A late time bottom-heavy IMF for massive present-day galaxies would help to explain the bimodal distribution of galaxies in another way. It would make the quenched galaxies effectively less quenched, because most of their star formation is in dwarf stars, the star formation rates implied by the observations would be higher. The requirements on the efficiency of the feedback process that is responsible for preventing gas accretion and for the ejection of gas lost by evolved stars would thus be reduced.

\begin{table*}
\begin{center}
\caption{The first column gives some typical stellar mass values for a universal Chabrier IMF. The second column gives the corresponding median stellar mass for a metallicity-dependent Vazdekis IMF, from Figure \ref{FigureMassShifts}. The third column gives the corresponding halo mass from \protect{ \citet{Behroozi10}}. The fourth column gives the median total HI plus H$\rm{_2}$ gas mass from \protect{\citet{Bahe15}} and references therein. The starred measurement indicates a slight extrapolation beyond $M_{\star}\approx11.3 {\rm M_{\odot}}$. There is at least a 0.5 dex uncertainty in this median gas mass value. The fifth and sixth columns give the median SFR for a Chabrier respectively Vazdekis IMF from Figure \ref{FigureMassSfr}. The seventh column gives the halo gas accretion rate at the given halo mass from \protect{\citet{Correa15}}. The accretion rates in the middle two lines are very close to the implied SFR for the Vazdekis IMF. The eighth and ninth columns give the median gas consumption timescales for respectively a Chabrier and Vazdekis IMF, calculated from the corresponding star formation rates and the hydrogen gas mass (column four), including a correction for helium. The Chabrier IMF implies a nearly constant median gas consumption timescale across 1.5 dex in mass, whereas the median gas consumption timescale for a Vazdekis IMF decreases by a factor 32 over this mass range.}
\begin{tabular}{llllllllll}
\hline
$log_{10}(M_{\star})$ & $log_{10}(M_{\star})$ & $log_{10}$ & $log_{10}$ & SFR & SFR & halo gas & gas consump- & gas consump- \\ 
Chabrier & Vazdekis & $(M_{Halo})$ & $(M_{HI+H_2})$ &  Chabrier & Vazdekis & accretion rate  & tion time & tion time \\
$\rm{[M_{\odot}]}$ & $\rm{[M_{\odot}]}$ & $\rm{[M_{\odot}]}$ & $\rm{[M_{\odot}]}$ & [$\rm{M_{\odot}/yr}$] & [$\rm{M_{\odot}/yr}$] & [$\rm{M_{\odot}/yr}$] & Chabrier [yr] & Vazdekis [yr]\\ \hline
10.0 & 10.3 & 11.6 & 9.2 & 1.0 & \;\;1.0 & \;\;2.5 & $\rm{2.1\times 10^9}$& $\rm{2.1\times 10^9}$\\  
10.5 & 10.8 & 12.1 & 9.5 & 1.8 & \;\;7.1 & \;\;8.0 & $\rm{2.3\times 10^9}$ & $\rm{5.9\times 10^8}$\\
11.0 & 11.3 & 13.1 & 9.5 & 2.5 & 50 & 79 & $\rm{1.7\times 10^9}$ &  $\rm{8.4\times 10^7}$  \\ 
11.5 & 11.8 & 14.9 & 9.4$^*$ & 2.5 & 50 &  $\rm{1\times 10^4}$& $\rm{1.3\times 10^9}$ & $\rm{6.7\times 10^7}$\\ \hline
\end{tabular}
\label{TableConsumption}
\end{center}
\end{table*}

The higher SFRs imply shorter gas consumption timescales. Previous work on the effect of a variable IMF on gas consumption timescales has been done by \citet{Pflamm09b} in the context of the IGIMF model. They report a gas consumption timescale of 3 Gyr, that does not depend on the neutral gas mass for dwarf irregular and large disk galaxies alike. Table \ref{TableConsumption} lists the median values for the gas consumption timescale implied by the SFR changes and the measurements of total HI plus H$\rm_2$ gas from the GASS and COLD GASS surveys \citep{Catinella10,Saintonge11}. For a universal Chabrier IMF the median gas consumption timescale is almost constant over the mass range from $\rm{10^{10}M_{\odot}}$ to $\rm{10^{11.5}M_{\odot}}$  at a value of $\rm{\approx2\times10^9}$ yr. Over the same mass range a metallicity-dependent Vazdekis IMF implies a progressively shorter median gas consumption timescale for more massive galaxies, with a decrease by a factor of 32 to a value of 67 million years for the most massive galaxies. If this star formation were to be sustained for longer periods, it would require an efficient refueling of the cool gas reservoir of 50 $\rm{M_{\odot}/yr}$ (in which case the refueling rate equals the star formation rate). We can compare this rate to the rate of gas accretion onto the halo. For this we use the stellar mass-halo mass relation from \citet{Behroozi10} to determine the halo mass (where we always use the Chabrier IMF) and the corresponding halo gas accretion rate from \citet{Correa15}. At a stellar mass of $\rm{10^{10} M_{\odot}}$, the accretion onto the halo is about a factor 2.5 larger than the median SFR, both for a universal Chabrier IMF and for the Vazdekis IMF. For the Chabrier IMF this ratio becomes progressively larger for larger masses, but for the Vazdekis IMF it first drops, around the knee of the galaxy mass function, where the halo gas accretion rate is just marginally larger than the SFR, before rising again at the high-mass end.

In conclusion, the increased SFR, relative to Chabrier, for the metallicity-dependent Vazdekis IMF may require large cooling flows onto the galaxy, of the same order as the gas accretion rate onto the halo, and would imply very short gas consumption timescales. The large gas accretion rates may be inconsistent with X-ray observations of galaxy clusters \citep[e.g.][]{Peterson03}. Combining the high-end metallicity-dependent IMF changes with changes to the low-end slope would reduce the magnitude of these effects. They would imply a speedup of the quenching process, especially around the knee of the galaxy stellar mass function.

\section{Conclusions}
\label{SectionConclusions}

In this paper we have investigated the consequences of a metallicity-dependent IMF, as measured for SDSS galaxies and elliptical annuli of CALIFA galaxies by \citet{MartinNavarro15}. We use the stellar mass and SFR measurements of \citet{Chang15} and the stellar metallicity measurements of \citet{Gallazzi05}. We assume that each of our 186,886 SDSS galaxies follows the exact IMF-metallicity relation given by equation \ref{eqone} \citep{MartinNavarro15}.

The measurement of IMF-specific spectral lines in ETGs by \citet{MartinNavarro15} is effectively sensitive to the current dwarf-to-giant ratio, equation \ref{eqfhalfone}. It is insensitive to the IMF slope for stellar masses $M\gg{\rm 1.0 M_{\odot}}$, because of a sparseness of stars in this mass range in the observed ETGs and because of the relative insensitivity of the used line indices to stars in this mass range. The implications of this varying IMF for the masses and star formation rates of observed galaxies are however very dependent on the high-end IMF. For this reason we investigate two extreme cases. First, we use the IMF parametrisation of \citet{MartinNavarro15}, for which the high-end IMF slope is varied and the low-end slope is held fixed. We refer to this case as the `Vazdekis IMF'. Second, we use a `low-end' IMF parametrisation, for which the high-end slope is fixed at the Chabrier value and the slope below 0.5 $\rm{M_{\odot}}$ is varied.  Our main conclusions are as follows:

\begin{itemize}
\item When we match both IMF parametrisations on the current dwarf-to-giant ratio, the same observations lead to completely different IMFs (Fig. \ref{FigureParametrisation}). 
\item The trend between the high-end IMF slope and metallicity can imply large changes in the SFR inferred from observations, depending on the SFR tracer. Compared to a Chabrier IMF, the implied bottom-heavy SFR can be up to three orders of magnitude larger for the most metal-rich galaxies, with a calibration offset between the observed FUV-SFR and the H$\rm{\alpha}$-SFR of up to an order of magnitude. The top-heavy IMF for metal-poor galaxies can reduce the inferred SFR by up to 0.8 dex. When we vary the low-end IMF slope instead of the high-end slope, the same IMF-metallicity trend causes much smaller SFR changes ranging from -0.02 dex to 0.20 dex and no FUV-H$\rm{\alpha}$ SFR calibration offsets (Fig. \ref{FigureSFR}).
\item The stellar mass changes implied by the IMF-metallicity trend are also much smaller when the low-end IMF slope is varied (-0.06 dex $<\Delta log_{10}(M_{\star})<$ 0.29 dex) than when the high-end IMF slope is varied (-0.06 dex $<\Delta log_{10}(M_{\star})<$ 1.25 dex). A combination of a metallicity-dependent variation of the low-end and high-end IMF slope would cause a behaviour that is in between these two extremes (Fig.  \ref{FigureMassChange}).
\item The high-end slope of 1.3 for a Chabrier IMF (or a similar Kroupa/Canonical IMF) minimizes the total mass of old galaxies. Changing the high-end slope will always increase the inferred mass of the galaxy by either locking up more mass in dwarf stars or in stellar remnants (Fig. \ref{FigureMassChange}).
\item The implied combined mass-SFR changes for our sample of 186,886 SDSS galaxies are shown in Figure \ref{FigureMassDifSfrDif}. When the high-end IMF slope is varied, metal-rich galaxies will become more massive and their SSFR will increase by up to two orders of magnitude. When changing the low-end IMF slope, the increase in mass is combined with a slight decrease in the SSFR.
\item Figure \ref{FigureMassSfr} shows the effect of a variable IMF on the star forming main sequence. The effect for the low-end IMF is mild, because most galaxies are displaced along the sequence. When changing the high-end slope the effect is large, especially at the high-mass end. The star-forming main sequence is effectively prolonged up to a stellar mass of $\rm{10^{12} M_{\odot}}$. In between $\rm{10^{9} M_{\odot}}$ and $\rm{10^{10} M_{\odot}}$ the median SFR becomes lower, but the same does not hold for the average SFR, due to an increased scatter.
\item Increases in the total low-redshift SFR density range from a factor of 1.11 for the low-end IMF to a factor of 16.6 for the Vazdekis IMF, with a larger SFR contribution from galaxies around a mass of $\rm{10^{11.3} M_{\odot}}$ (Fig. \ref{FigureSfrDensity}).
\item If the metallicity-dependent IMF variations only affect the low-end slope, then they will produce a clear trend of the IMF-dependent mass change with SED-fitting mass (using a Chabrier IMF for the SED fit) (Fig. \ref{FigureMassShifts}). For $M\gtrsim {\rm10^{11} M_{\odot}}$ a small fraction of galaxies will be more massive than for a Salpeter IMF. High-end IMF slope variations produce a larger mass change, possibly in tension with dynamic mass measurements, but no clear trend with Chabrier SED mass.
\item Independent of the IMF parametrisation, the IMF-metallicity trend will produce a shift of the galaxy stellar mass function to higher masses, without changing the steepness of the high-mass dropoff much, see Figure \ref{FigureGSMF}. This means that these IMF variations do not remove the need for an efficient quenching mechanism for high-mass galaxies, but they shift the mass at which this happens upwards by 0.2 to 0.5 dex. The implied total stellar mass density of the Universe increases by a factor of 1.3 to 2.3.
\item The IMF-metallicity trend implies a steeper stellar mass-metallicity relation for $M\sim {\rm 10^{11} M_{\odot}}$ (Fig. \ref{FigureMassMetallicity}).
\item If the high-end IMF slope is more shallow for metal-poor galaxies, then this would produce a large change in the metal mass released per unit mass of stars formed over the lifetime of a galaxy. Figure \ref{FigureYields} shows that this metal production could be three orders of magnitude higher in the early metal-poor stage of a galaxy than in the late metal-rich stage. This strong evolution makes it impossible to give a good estimate of the total metal production of any galaxy as this might be dominated by the early stages and depends crucially on the exact mass and metallicity evolution of the galaxy. The total metal production deduced from the H$\rm{\alpha}$-SFR evolution of the Universe is not expected to change much though, since H$\rm{\alpha}$ directly traces the massive metal-producing stars.
\item If the high-end IMF slope is shallower for metal-poor galaxies, then this will increase the production of ionising photons for metal-poor galaxies in the early Universe. This would imply an increased production of ionising photons for high redshifts that are not probed directly by $\rm{H\alpha}$ observations, which would have an effect on models for the reionisation of the Universe.
\item A bottom-heavy IMF during the late, metal-rich stage of galaxy evolution could help to explain the observed bimodality in galaxy properties: the observation that most galaxies can be classified as `blue cloud' or `red sequence' galaxies and only very few galaxies are observed in the intermediate `green valley' stage. On the one hand, `quenched' galaxies would be less quenched, because their true star formation rate would be higher than for a Chabrier IMF, due to the `invisible' formation of dwarf stars. On the other hand, the quenching time of these galaxies would be lower, due to the much shorter implied gas consumption timescale. If gas cooling onto the galaxy is regulated in part by stellar feedback, then a more bottom-heavy IMF at higher metallicity could initiate a runaway process of low-mass star formation, because these low-mass stars do not provide feedback.
\item Metallicity-dependent changes to the high-end IMF slope imply gas consumption timescales that become progressively shorter for more massive galaxies, down to 67 million years at $\rm{10^{11.5}M_{\odot}}$, whereas for a universal Chabrier IMF this gas consumption timescale is approximately constant at $\rm{\approx 2 \times 10^9 yr}$. Around the knee of the GSMF these IMF changes require a cooling flow onto the galaxy of about 7 to 50 $\rm{M_{\odot}/yr}$ in order to sustain the SFR. These values are only slightly below the expected gas accretion rate onto the halo, see Table \ref{TableConsumption}.
\end{itemize}

In conclusion, the implications of the observational evidence for a variable IMF for the low-redshift Universe could vary from absolutely dramatic to mild but significant, depending on the assumed IMF parametrisation.

\section*{Acknowledgements}

We would like to thank Chris Barber, Ignacio Mart\'in-Navarro, Alexandre Vazdekis and Francesco La Barbera for valuable contributions. We gratefully acknowledge support from the European Research Council under the European Union's Seventh Framework Programme (FP7/2007-2013) / ERC Grant agreement 278594-GasAroundGalaxies.

\bibliographystyle{mn2e} 
\bibliography{Bibliography}

\appendix{}
\section{}

\begin{figure*}
\center
\includegraphics[height=0.9\columnwidth]{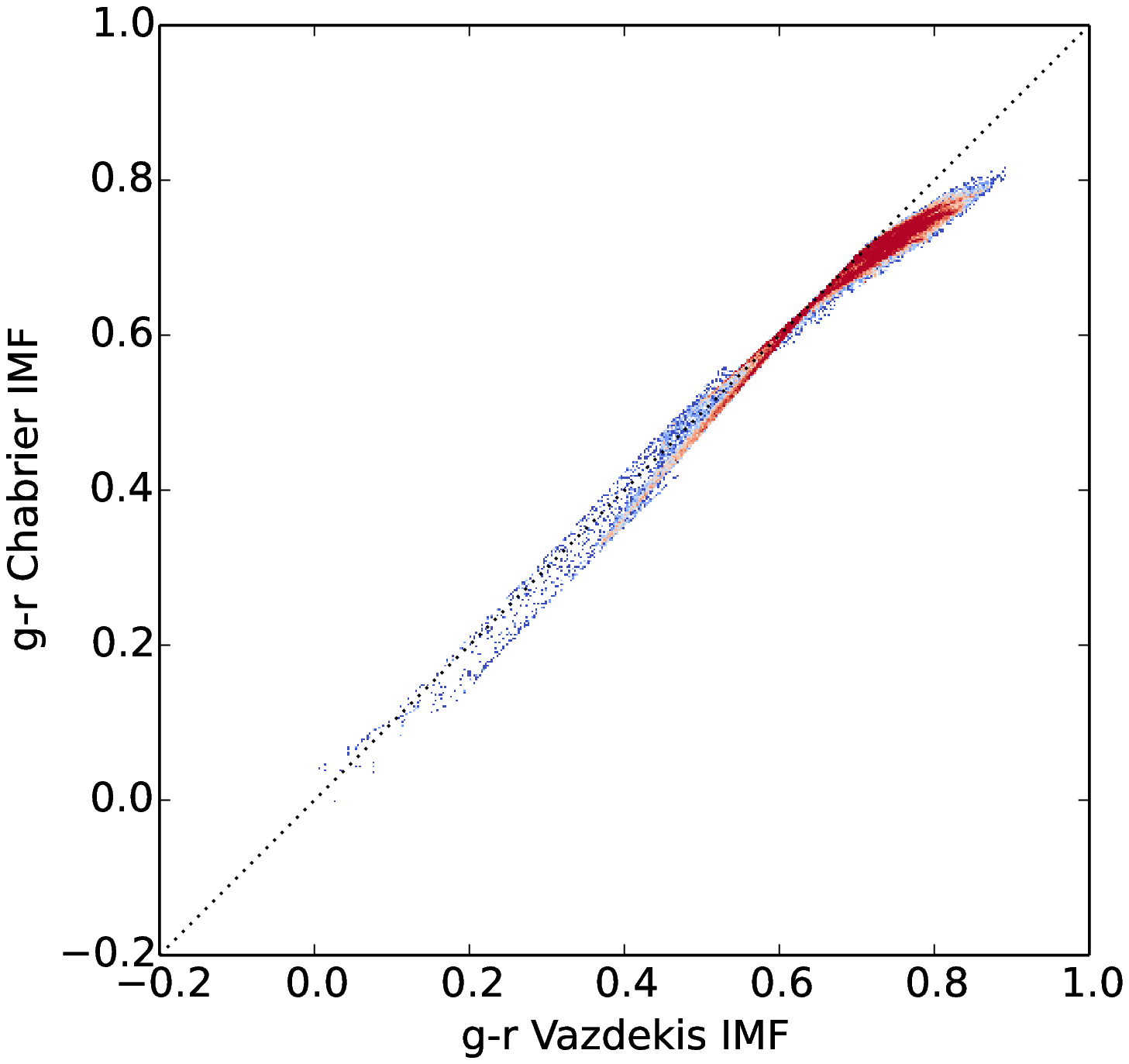}
\includegraphics[height=0.9\columnwidth]{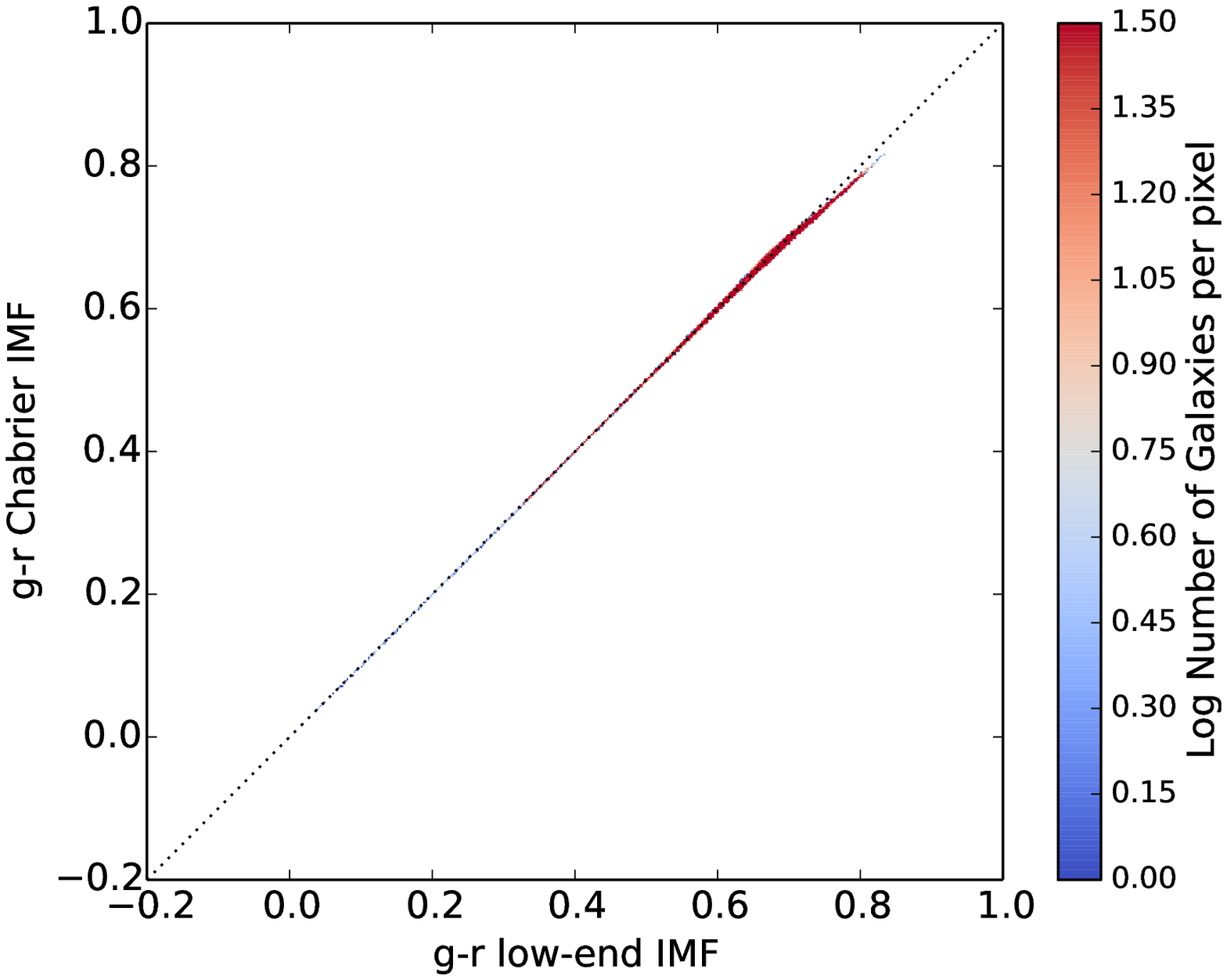}
\caption{Dustless SDSS g-r colour of the varying IMF versus the colour of a Chabrier IMF for the complete sample of galaxies used in this work. The left panel show the Vazdekis IMF and the right panel shows the low-end IMF. The black dotted line indicates a one-to-one relation. A large deviation from this would mean a breakdown of our method to recalibrate stellar masses to a changed IMF. In the `low-end' case the colours are almost unchanged, because the dwarf stars do not contribute much to the overall light. Also in the `Vazdekis' case the colour changes are quite small.}
\label{FigureColours}
\end{figure*}

In section \ref{SectionMethod} we described the method by which we calculate new masses and star formation rates for galaxies with an IMF different than the one used in the original mass and SFR determination. The most precise way to do this would be by redoing the whole SED fitting procedure for every galaxy, assuming a metallicity-dependent IMF. In this work we have used a simpler method to  recalibrate the stellar masses and star formation rates, because at the moment we believe the errors to be dominated by uncertainties in the IMF measurements and in the stellar metallicity measurements and by a large dependency on the assumed IMF parametrisation. As a first order correction to the masses we compare mass-to-light ratios in the Cousins R-band for a simple stellar population with an r-band weighted age and metallicity given by \citet{Gallazzi05} and a varying IMF to the mass-to-light ratio of a SSP with the same age and metallicity, but a Chabrier IMF. An alternative method would be to match on colour and some combination of age and metallicity. The problem with matching on colour is that it is very susceptible to dust. Still, we expect our method to break down once the dustless colour of the varied IMF SSP is very different from that of the Chabrier SSP, because a large change in dustless colour would also imply a large change in the true observed colour.

For this reason we plot the dustless SDSS g-r colours of our complete galaxy sample in figure \ref{FigureColours}. For the low-end IMF parametrisation there is almost no change in colour. This parametrisation only changes the IMF below 0.5 $\rm{M_{\odot}}$. These stars do not contribute much to the overall light and as a consequence changing the low-end IMF has a negligeable effect on the dustless colour. Also for the Vazdekis IMF parametrisation the changes in dustless colours are quite small.

\begin{figure}
\center
\includegraphics[width=1.0\columnwidth]{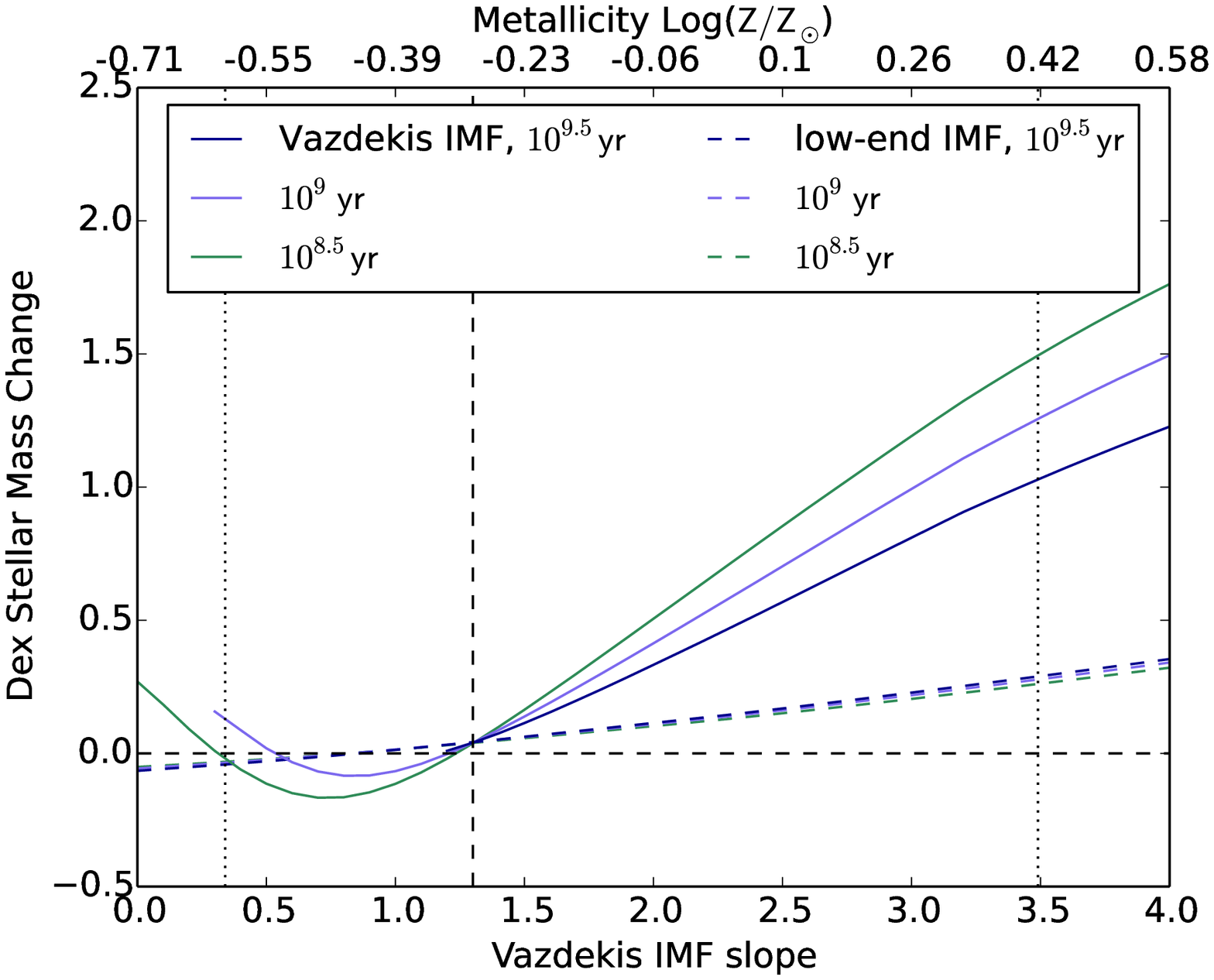}
\caption{Same as figure \ref{FigureMassChange}, but for a constant star formation history instead of a simple stellar population. This plot is meant to estimate the error made in using a SSP for all galaxies. The different colours denote different r-band weighted ages. The solid lines stop at the left when the implied start of constant star formation would be the same as the age of the Universe.}
\label{FigureMassChangeB}
\end{figure}

Another simplification lies in the fact that we use a simple stellar population, which is a good approximation for ETGs but not for star forming galaxies. For a star forming galaxy the used SSP is a simplification with respect to a constant star formation history or an exponential one, but the used r-band luminosity weighted age is skewed towards more recent star formation, as opposed to a mass-weighted age, through a large dependency of luminosity on mass. Typically the r-band luminosity weighted age is about half of the true average stellar age for a Chabrier constant star formation rate older than $\rm{10^{8} yr}$. In order to estimate the error in the IMF dependent mass change introduced by our simplification of the star formation history we plot the expected mass changes for a constant star formation history in Figure \ref{FigureMassChangeB}. This should be compared with Figure \ref{FigureMassChange}. The labeled ages are r-band luminosity weighted ages. As a function of IMF slope also the ratio between the r-band weighted age and the age of the start of star formation changes. This ratio becomes increasingly larger for an increasingly top-heavy IMF, which causes the cut-off of the Vazdekis IMF lines at the left. Here the implied age to the start of the constant star formation would be more than the age of the universe. The assumption of a SSP has no drastic consequences. For a low-end IMF parametrisation the results are almost indistinguishable. For a Vazdekis IMF parametrisation with a super-Chabrier slope, a constant star formation rate increases the expected dex mass shifts by 15\%-25\%. For a sub-Chabrier IMF slope the effect is more complicated. Here, a constant star formation history is only possible for low r-band weighted ages. The implied mass changes will be up to 0.15 dex lower than for a SSP. In some cases this may make the galaxy lighter than Chabrier instead of heavier, with a lightest mass change of -0.17 dex.

We conclude that our method is a good first order approximation of the expected mass differences implied by a metallicity-dependent IMF. A more precise treatment, using a varying IMF in SED fitting could be beneficial, but at the moment this is not expected to be the main source of systematic or random errors. Moreover SED fitting itself also suffers from an age-metallicity degeneracy and a degeneracy in the fitted star formation history. For example \citet{Torrey15} find that SED fitting mock galaxies from the Illustris simulation with the FAST code delivers more accurate stellar masses when the metallicity for all galaxies is kept fixed.

\label{lastpage}
\end{document}